Nanoporous High Entropy Alloys: Overcoming Brittleness Through Strain Hardening

Author names and affiliations:
Jarod Worden[a], Juergen Biener[c], Celine Hin[a,b],

a. ME Department
Northern Virginia Center, Virginia Polytechnic Institute
7054 Haycock Rd, Falls Church, VA 22043, United States
b. MSE Department
Northern Virginia Center, Virginia Polytechnic Institute
7054 Haycock Rd, Falls Church, VA 22043, United States
c. Materials Science Division
Physical and Life Sciences Directorate, Lawrence Livermore National Laboratory
Livermore, CA 94550, United States

Corresponding Authors
E-mail addresses: celhin@vt.edu (Celine Hin); jaworden@vt.edu (Jarod Worden),
Postal Address: Northern Virginia Center, Virginia Polytechnic Institute
7054 Haycock Rd, Falls Church, VA 22043, United States

**Abstract**

Bicontinuous nanoporous materials possess remarkable mechanical properties, such as higher specific strength and lower specific modulus compared to fully dense materials combined with low densities and high specific surface areas. Unfortunately, their practical application is hindered by inherent macroscopic brittleness, mainly due to cascading ligament failure under tension. To address this limitation, we investigate whether high entropy alloys, recognized for their outstanding strength and strain hardening properties, can mitigate nanoporous material's inherent brittleness. Molecular dynamics simulations of nanoporous $Al_{0.1}CoCrFeNi$ and NbMoTaW reveal a dual mechanism involving dislocation starvation and sluggish dislocation motion, resulting in specific strength values 5 to 10 times higher than those of single-element nanoporous materials, and a resilience against thermal degradation. Strain hardening, driven by sluggish dislocations, effectively prevents failure of the weakest ligaments under tensile stress in face-centered cubic architectures by trapping stacking faults in the ligaments and dislocation forest hardening in the nodes of body-centered cubic structures, demonstrating their potential to shape the next generation of high strength, low density materials.

## 1. Introduction

Nanoporous metals, featuring random bicontinuous ligament-pore structures, possess unique mechanical properties, including high specific strength, good ductility, and energy absorption, all while maintaining lower densities compared to typical bulk metals [1,2]. However, their practical application is hindered by brittleness under tension, as fracturing of the weakest ligaments can lead to a cascading fracture of adjacent ligaments [3,4]. For example, nanoporous gold, studied extensively through simulations and experiments, shows macroscopic brittle behavior triggered by single ligament failure [5,6].

To address the brittleness challenge associated with cascading ligament failure, we investigate High Entropy Alloys (HEAs), known for substantial strain hardening, in nanoporous structures [7,8]. HEAs use equiatomic concentration of multiple elements to increase system entropy, stabilizing disordered solid-solution phases and preventing brittle intermetallic phases [1,9]. They offer high strength, hardness, strain hardening, and resistance to fatigue, creep, and wear, overcoming challenges in single-element nanoporous materials [9,10]. Our hypothesis is that incorporating HEAs into bicontinuous nanoporous structures will enhance mechanical properties like strength and toughness while maintaining low densities compared to traditional metals and alloys. This makes them promising for applications requiring superior mechanical performance combined with low densities and high thermal tolerance, benefiting industries such as nuclear, automotive, aerospace, and aeronautics compared to commonly used industrial materials [11,12] (further investigation of these materials in automotive and nuclear are explained in Supplementary Discussion 1). In the realm of high energy absorption and impact resistance, nanoporous HEAs are expected to showcase appealing mechanical properties that optimize energy deflection and allowable stress [13]. However, their application is hindered due to their novelty and lack of mechanical experimental and computational data.

This study employs Molecular Dynamics (MD) simulations to investigate the mechanical behavior of nanoporous High Entropy Alloys (HEAs) under uniaxial compression and tensile testing. For comparison, we also performed simulations on fully dense single crystalline (SC) and polycrystalline HEAs. The selection of $Al_{0.1}CoCrFeNi$ and NbMoTaW HEAs is based on their distinctive single disordered face-centered cubic (fcc) and body-centered cubic (bcc) phase [14]. The simulation outcomes validate our hypothesis that nanoporous HEAs exhibit unique mechanical properties, especially significantly higher specific strength compared to single-element nanoporous materials (Figure 1). The high specific strength can be attributed to the sluggish dislocation mechanism observed in the studied HEAs that enhances the strength of individual ligaments by enabling strain hardening mechanisms such as dislocation forest hardening, effectively addressing the macroscopic brittleness associated with nanoporous materials.

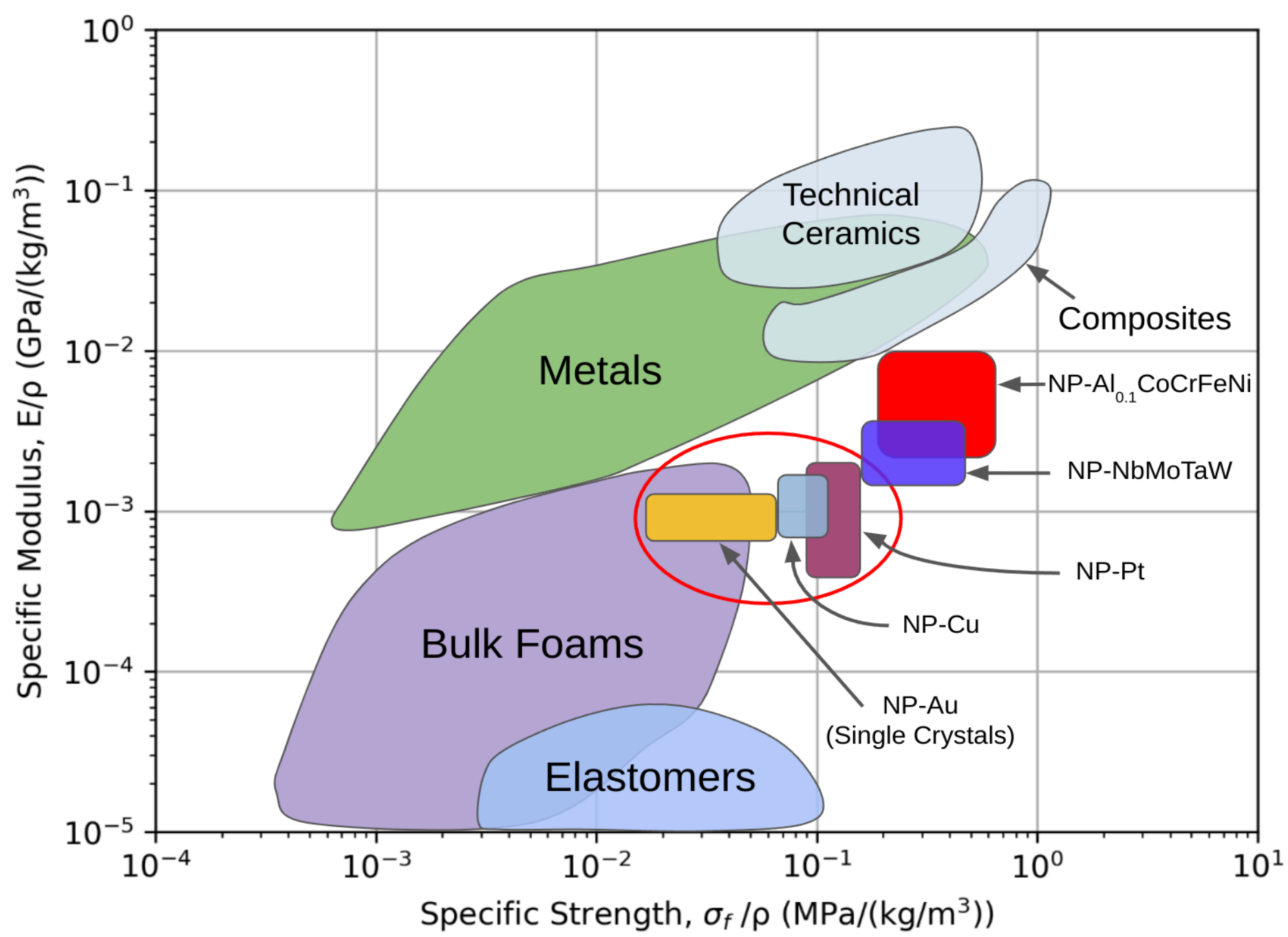

**Figure 1.** Weight-normalized Young's modulus and yield strength for different classes of materials including bulk materials and nanoporous materials [15]. The red circle denotes the general phase space occupied by single-component nanoporous materials based on available data. The red box represents the nanoporous HEA $Al_{0.1}CoCrFeNi$ and the blue box represents the nanoporous HEA NbMoTaW, highlighting their distinctive values in specific strength and specific modulus.

## 2. Methods

The molecular dynamics (MD) simulations were conducted using the open-source code LAMMPS, employing a three-dimensional cell with periodic boundary conditions, with dimensions of 44.625 nm x 44.625 nm x 44.625 nm [16]. The bulk cell was generated by populating it with 7,812,500 fcc and 3,906,250 bcc structural atoms. The concentrations of Al, Co, Cr, Fe, and Ni (or Nb, Mo, Ta, W) were randomly assigned using Atomsk [17]. Polycrystalline systems were created by applying Voronoi Tessellation to the bulk cell with random grain orientations. Three distinct polycrystalline structures with grain sizes of 10 nm, 20 nm, and 30 nm were generated. The methodology for producing nanoporous HEAs followed a similar approach to that used for simulating the creation of bicontinuous nanoporous gold through dealloying methods and are described in Supplementary Discussion 2 [18]. The types of systems in this study are depicted in Figure 2.

$Al_{0.1}CoCrFeNi$ system was simulated using an Embedded Atom Method (EAM) potential, which has also been employed in previous MD investigations to study nanoindentation tests and plastic deformation [19–21]. The stacking fault energy (SFE) of the single-crystal (SC) system was calculated to be 32.8 mJ/m$^2$, showing good agreement with the experimental value of 30 mJ/m$^2$ and aligning with other computational findings [22,23]. This low stacking fault energy of the potential is similar to other fcc HEAs such as the cantor alloy CoCrFeMnNi where the stacking fault energy is also found to be exceedingly low (< 35 $mJ/m^2$) using both experimental and computational methods, much lower than that of single element materials [24,25] .

For the NbMoTaW system, a Spectral Neighbor Analysis Potential (SNAP) potential was applied, which is capable of accurately reproducing elastic constants [26] . To analyze defects within the system, the Dislocation Analysis Tool was employed to assess dislocation densities, dislocation types, and the formation of Stacking Faults/Twin Boundaries [27]. The ligament diameter measurement involved identifying specific atom positions as the center of an expanding sphere. This sphere grew radially by half the lattice parameter. We assessed if a part of the sphere contacted the ligament's surface by examining interactions between the sphere and local atom configuration. If contact was confirmed on one side, we checked the opposite side. When both sides contacted the surface, we calculated the distance as the ligament section's diameter, contributing to the average ligament diameter calculation for the entire system. Average ligament size is displayed in Table 1 with error.

Molecular dynamics simulations were employed to conduct mechanical compression and tensile tests at temperatures of 298 K, 600 K, and 1273 K. Initially, a Nose/Hoover thermostat and barostat (NPT) ensemble was utilized for a thermalization period of 100.0 ps. This thermalization phase took place at the specified temperatures and atmospheric pressure until the system achieved a stable state with respect to lattice parameters and energy. Following the thermalization process, the NPT ensemble was re-employed, and a strain rate of $10^9$ $s^{-1}$ was applied to deform the simulation box along the x direction. Compression tests were conducted in the range of 0-70% strain, while tensile tests covered the 0-50% strain range.

The use of high strain rates in MD simulations is a recognized limitation, as realistically slow strain rates require prohibitively long simulation times. High strain rates generally result in higher ultimate stresses, making direct comparisons with experimental data more challenging. Our study, along with others in the field, examines various strain rates and finds that differences between lower strain rates are relatively minor, indicating a diminishing return in the reduction of ultimate stress with decreasing strain rate [28,29]. This trend is highlighted in Supplementary Table 1, which presents ultimate stress calculations for different tensile strain rates in the 28.0% relative density nanoporous NbMoTaW system. Notably, at a lower strain rate, a minimal decrease in ultimate stress is observed (0.02 GPa), whereas a significantly higher stress is recorded at an increased strain rate (0.12 GPa). The chosen strain rate effectively balances the need for faster simulation times while minimizing its impact on stress results.

## 3. Results

The results section analyzes the mechanical properties involving the HEA systems through stress-strain curve analysis, ligament fracture analysis, defect analysis, and surface energy analysis. The analysis focuses on random placement of atomic species for the different HEA systems, however there is the potential for surface segregation due to the relatively large surface area. This can be analyzed in future investigations.

### 3.1 *Stress-Strain curve analysis*

Figure 3(a-d) displays stress vs. strain curves at 600 K (more results from tests at 298/600/1273 K are shown in Figure S1 and Figure S2) for $Al_{0.1}CoCrFeNi$ and NbMoTaW while the corresponding yield strength and Young's modulus data are summarized in Tables 2 to 5. The highest strain values displayed for the nanoporous systems under compression correspond to strains before full densification, which occurs at approximately 70%, 60%, and 50% strain for the 28.9%, 47.8%, and 67.2% porosity samples, respectively (Fig. 3).

For both compression and tension, and irrespective of temperature, the $Al_{0.1}CoCrFeNi$ systems exhibit linear elastic behavior within the low strain range up to 2%. For non-nanoporous structures, a transition to a non-linear elastic deformation behavior can be observed with further increasing strain until the ultimate stress point is reached at 5% strain. In the case of SC systems, the yield strength is attained early, in contrast to the ultimate strength, which is achieved at higher strains, indicating the strain-hardening mechanism of the HEAs [10]. This is observed in the compressive yield strength displayed in Table 2. For the nanocrystalline (NC) systems, both under compression and tension, the stress drops significantly after reaching the yield peak, followed by a plateau region that continues up to the highest strain values. Notably, in the NC systems, compression tests yield higher stress values compared to tensile tests (Figure 3a). This observed compression-tension asymmetry in yield strength for SC and polycrystalline $Al_{0.1}CoCrFeNi$ has also been reported in other HEA systems [30], possibly reflecting distinct deformation mechanisms such as twinning in compression and slip in tension [31].

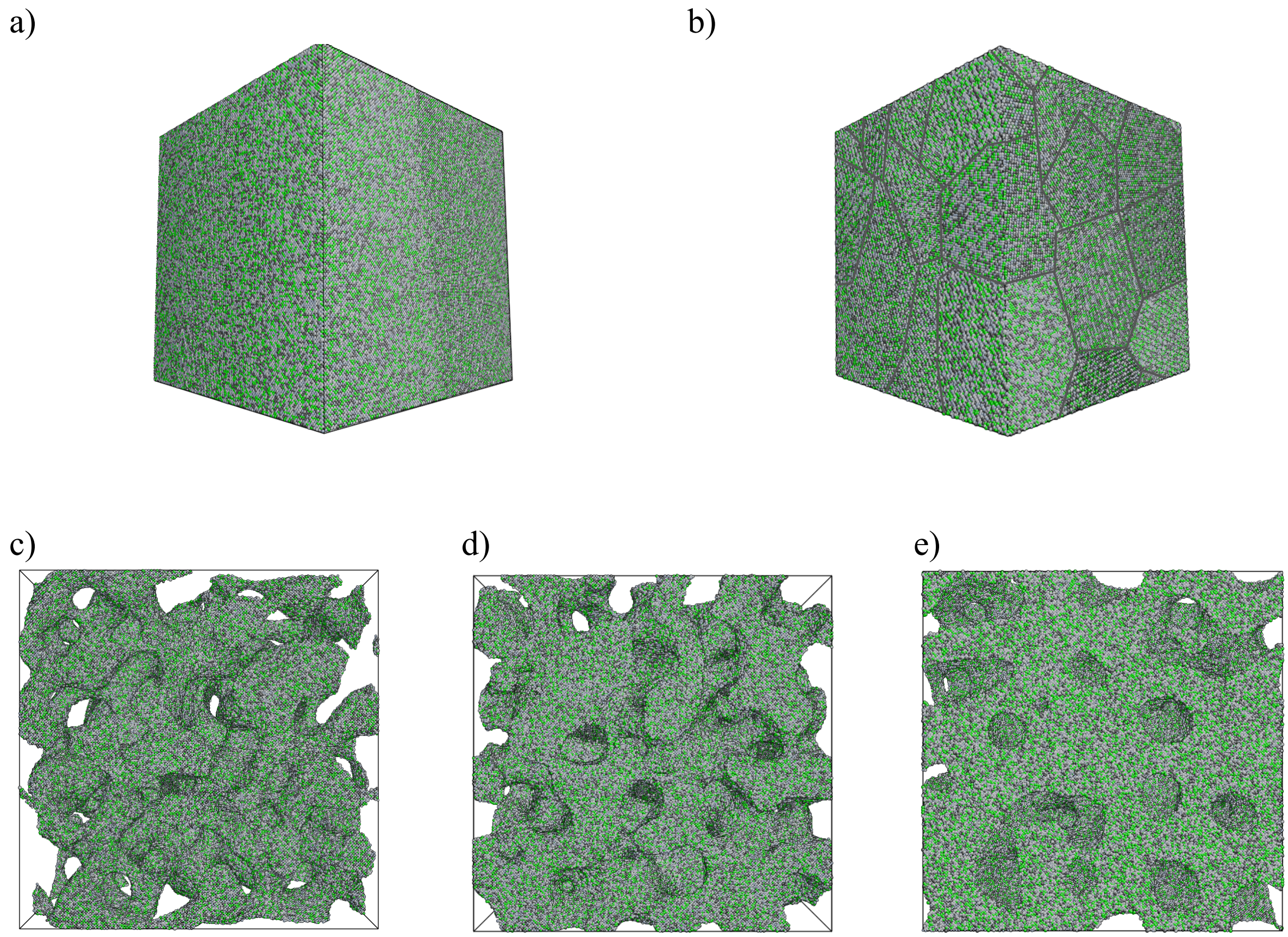

**Figure 2.** The simulation boxes of the HEA $Al_{0.1}CoCrFeNi$ systems for (a) single crystalline, (b) 30 nm polycrystalline, and nanoporous $Al_{0.1}CoCrFeNi$ with relative densities of (c) 28.7, (d) 47.8, and (e) 67.2%.

The stress-strain response of nanoporous $Al_{0.1}CoCrFeNi$ under both compression and tensile test conditions is more complex when compared to SC and NC systems (Fig. 3a). Tensile tests conducted on nanoporous $Al_{0.1}CoCrFeNi$ exhibit a more distinct yield peak, followed by a shorter plateau phase, before the stress decreases with further increasing strain. These trends align with those observed in other FeCrNiCuCo fcc HEA nanoporous systems [28,32]. In compression tests, only a small yield peak was observed for the highest relative densities, followed by an immediate stress plateau and a subsequent region where stress gradually increases with strain. This response aligns with what is commonly observed in many nanoporous systems under uniaxial compression in both experimental and simulated settings for single element and HEA systems [33,34]. Comparing the effect of relative density and ligament size on compression stress-strain curves (Figure 3b) reveals that at any given strain within the plateau region, higher relative density correlates with higher stress. Only the 67.2% relative density / 7.33 nm ligament diameter sample shows a distinct yield peak under compression; with increasing strain the effect of ligament size on the stress response becomes smaller and smaller.

This mirrors patterns observed in FeCrNiCuCo nanoporous fcc HEA systems, where it was found that relative density exerted a more significant influence on mechanical properties compared to ligament size, particularly when the ligament size is very small (Fig. 3b) [28].

The mechanical properties of $Al_{0.1}CoCrFeNi$ systems were also found to be temperature-dependent, with elevated temperatures resulting in reduced stresses across all systems (Fig. 3c and Fig. S1). The most pronounced stress reduction with temperature occurs in the 28.9% relative density sample. Conversely, the SC sample shows minimal or even reversed temperature dependence beyond the yield point. Additionally, for systems with similar relative densities, smaller ligaments lead to a more pronounced reduction in stress under compression at high temperatures (1273 K vs 298 K and 600 K) compared to larger ligament systems at higher temperatures, as indicated in Tables 1-4. This effect is particularly evident for the 50% relative density samples, as shown in Fig. S1a-c. The softening of smaller ligament systems with increasing temperatures may reflect the limited thermal stability of the smallest ligaments, as evidenced by the collapse of the ligament structure in the 28.7% relative density system with 2.2 nm ligaments during the initial relaxation phase. Under tensile test conditions, the np-$Al_{0.1}CoCrFeNi$ samples exhibit increased plasticity with increasing temperature (Fig.S1d-f).

The compression behavior of the NbMoTaW systems (as shown in Figure 3d) followed similar stress-strain trends as the $Al_{0.1}CoCrFeNi$ systems, including an initial steep increase in stress and a more or less pronounced yield peak at around 5% strain. This is followed by a plateau region where stress remain relatively constant with increasing strain. In the case of tensile strain (Figure S2 in Supplementary Materials), the SC NbMoTaW system displayed stress oscillations after the failure strain, which corresponded to oscillating atomic positions of the atoms following system rupture. Increasing temperature also resulted in decreased stress across all systems, with a more significant reduction observed for SC and NC structures.

Overall, the NbMoTaW systems exhibited superior mechanical properties, including higher yield strength and Young's modulus, compared to the $Al_{0.1}CoCrFeNi$ systems (Figure 3d and S2). An exception is observed in the 20 nm NC samples, where NbMoTaW shows lower stress than $Al_{0.1}CoCrFeNi$. This may reflect differences in defect interactions or simulation setup and warrants further investigation. The higher yield strength and Young's modulus of the NbMoTaW samples can be attributed to the intrinsic properties of the individual elements of the NbMoTaW system, which inherently possess greater yield strength, ultimate strength, and Young's modulus in both compressive and tensile environments than the elements in the $Al_{0.1}CoCrFeNi$ [35].

Tables 1 and 2 present the yield stress and Young's modulus for the $Al_{0.1}CoCrFeNi$ systems at both 298 K and 1273 K. Notably, the Young's modulus obtained from tensile mechanical tests for $Al_{0.1}CoCrFeNi$ closely aligns with experimental data and is consistent with previous molecular dynamics simulation results [22,23,36,37]. Additionally, the yield stress and strain results obtained from both tensile and compression tests show strong agreement with other molecular dynamics studies [38,39]. Tables 3 and 4 in the main text provide the yield stress and Young's modulus for NbMoTaW. The compressive modulus also closely corresponds with experimental data and aligns with other simulation results obtained under compression conditions [40–42], further supporting the reliability of the interatomic potential used.

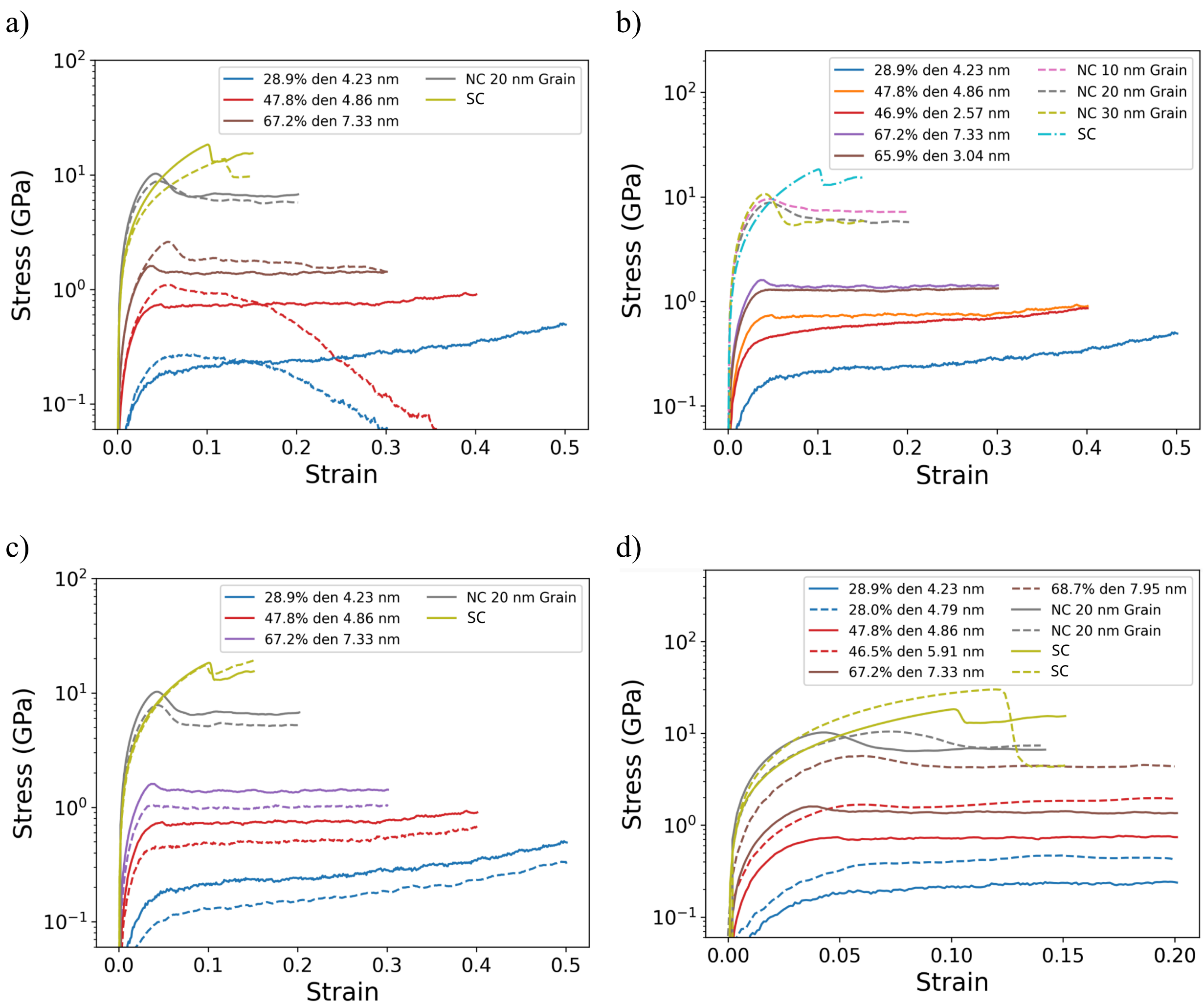

**Figure 3.** Stress-strain behavior of the studied HEA systems: (a) Compression (solid line) vs tension (dashed line) stress-strain curves obtained from nanoporous $Al_{0.1}CoCrFeNi$ samples at 600 K compared to fully dense SC and NC (20 nm) test samples; (b) Effect of ligament diameter and density of nanoporous $Al_{0.1}CoCrFeNi$ samples on compressive stress-strain behavior, (c) Effect of temperature on compression behavior of $Al_{0.1}CoCrFeNi$ : 600 K (solid line) vs. 1273 K (dashed line), (d) Comparison of compression behavior of $Al_{0.1}CoCrFeNi$ (solid line) and NbMoTaW (dashed line) at 600 K.

Additionally, the nanoporous systems exhibit a tension-compression asymmetry (see Table 2-5), a phenomenon previously observed in other nanoporous systems where compression demonstrates higher yield strength and Young's modulus [33]. This asymmetry varies depending on the type of high entropy alloy and crystal structure. For instance, the $Al_{0.1}CoCrFeNi$ nanoporous structure exhibits a higher compression yield strength compared to tension, while the NbMoTaW nanoporous system displays a higher yield strength under compression conditions. This discrepancy can be attributed to distinct defect mechanisms involved in plastic deformation. In bcc structures, common defect mechanisms include twinning and anti-twinning on the {211} slip planes, as well as non-planar core structures of screw dislocations [42]. On the other hand, in fcc structures, the asymmetry may be due to biased surface stresses and differing slip planes during tensile and compressive yielding [43].

### 3.2 Fracture Analysis in Tensile Tests of Nanoporous $Al_{0.1}CoCrFeNi$

Fracture analysis was performed exclusively on specimens subjected to tensile testing as no fractures were observed during compression tests due to predominant ligament bending [44]. This analysis focused on a segment of the nanoporous $Al_{0.1}CoCrFeNi$ system with a relative density of 28.9% and ligament size of 4.23 nm, tested under tensile conditions at 1273 K. Surface topography, dislocations, twin boundaries, and stacking faults were analyzed using the Dislocation Extraction Algorithm (DXA) and Paraview, as depicted in Figure 4a-l [45,46].

Dislocations are initially present during the early stages of fracture but gradually diminish as ligament failure progresses (Figure 4d,e,f). Twin boundaries form at the onset of fracture and persist throughout ligament failure; however, they disintegrate as the ligament ruptures (Figure 4g,h,i). Stacking faults are the most prevalent defects observed, remaining dominant in the ligament at all stages of the fracture process (Figure 4j,k,l). This behavior can be attributed to the low stacking fault energy characteristic of the HEA system.

This distinguishes the fracture processes of HEA ligaments from those observed in ligaments of nanoporous gold where dislocations appear to play a pivotal role in the ligament deformation [6,36,37]. The absence of distinct dislocations in the HEA ligaments during fracturing may be attributed to two factors: firstly, the hindrance of dislocation motion due to lattice distortion caused by the different atomic species in the HEA material, which is expected to reduce the formation of larger dislocations; and secondly, the high density of stacking faults impeding dislocations [36]. The sluggish dislocation motion of the HEA is corroborated by investigating the higher Potential Energy Landscape (PEL) of the ½<110>{111} edge dislocation which has higher energy barriers compared to single element PELs as shown in Supplementary Figure S3 and discussed in Supplementary Discussion 3. Given that the primary dislocation type is partial dislocations, neighboring partial dislocations formed from dissociated edge dislocations in the ligament create highly stable stacking faults between them, stemming from impaired dislocation mobility and low SFE in the nanoporous HEA system. As a result, the number of independent dislocations decreases, and the primary mechanism for dislocation elimination in the ligaments is the breaking of stacking faults [38]. The presence of stacking faults and partial dislocations within ligaments seems to be a prevalent mechanism in FeCrNiCuCo fcc HEA nanoporous structures, suggesting potential similarities in defect responses across these materials [28,32]. As dislocation motion typically constitutes a primary mechanism for plastic deformation, the absence of dislocations implies the existence of an alternative process for ligament fracturing during uniaxial tension [39,40]. Owing to the substantial prevalence of stacking faults in the ligament structure during fracture, the principal process bears a close resemblance to the twinning mechanism observed in fcc metallic materials, which is particularly pronounced during severe plastic deformation [41].

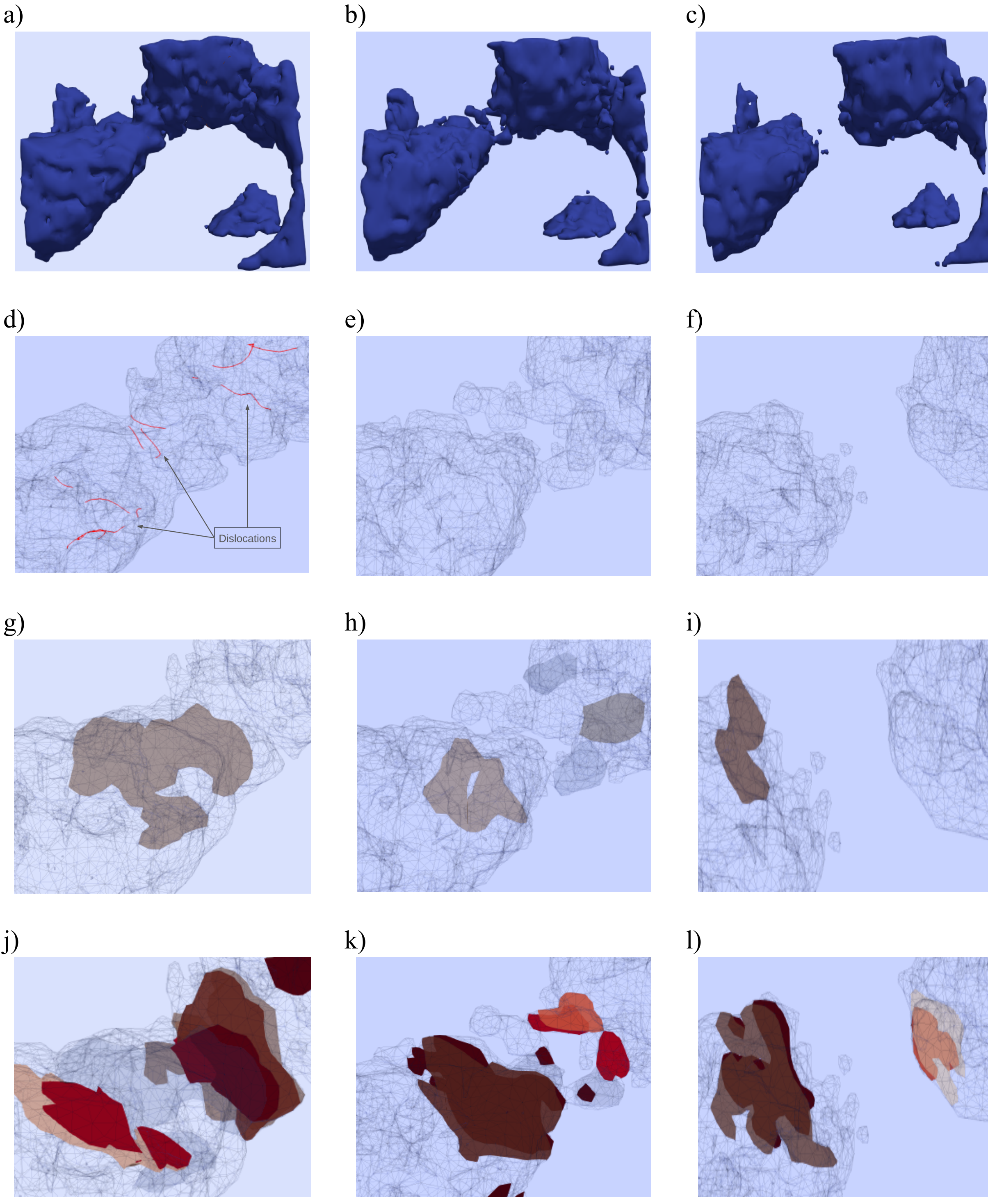

**Figure 4.** Snapshots of the fracture process of a 4.29 nm ligament from a nanoporous $Al_{0.1}CoCrFeNi$ HEA sample (28.9% relative density) under tensile testing: (a,b,c) surface solid mesh outlining the atomic surface, (d,e,f) dislocations (marked in red) in the ligament, (g,h,i) twin boundaries, and (j,k,l) stacking faults. The strain values represented in the first, second, and third column are 22.5%, 26.25%, and 31.25%, respectively.

### 3.3 Fracture Analysis in Tensile Tests of Nanoporous NbMoTaW

To investigate defects in the bcc structure, the bcc defect analysis (BDA) tool is used instead of the DXA tool, as justified in Supplementary Discussion 4 with accompanying Supplementary Figure S9 [47]. Below 14% strain, no defects were observed within the ligaments; however, at higher strain levels, twin boundaries nucleate alongside dislocations concentrated around the edges (Figure 5a). While the BDA tool does not classify dislocation types, it is important to note that bcc systems commonly exhibit a/6<111> partial dislocations, which play a crucial role in twinning - a key deformation mechanism in bcc Fe nanopillars [42]. These partial dislocations facilitate the movement of twin boundaries toward the ligament's center (Figure 5b), where they merge into complex dislocation bundles (Figure 5c). This interaction is essential for atom removal at the ligament's notch, ultimately leading to fracture (Figure 5d)

### 3.4 Defect Analysis in Nanoporous Systems

The effects of test temperature and geometry on dislocation and stacking fault concentrations in the $Al_{0.1}CoCrFeNi$ system are summarized in Figure 6, while total defect densities are presented in Figures S4 and S5. In both geometries, the dislocation density of nanoporous $Al_{0.1}CoCrFeNi$ (Figure 6a,c,e) exhibits a sharp increase up to 10% strain, followed by either a plateau (tensile tests) or a continuous increase (compression tests) for higher strains. The stacking fault density of the nanoporous systems exhibits a steadier growth that is also more influenced by temperature and testing type (Figure 6b,d,f).

Higher temperatures generally lead to a reduction in defect density due to faster dislocation annihilation and elevated stacking fault energy (Figure 6c and 6d) [43,48]. Defect annihilation is further facilitated by the higher surface area-to-volume ratio in lower relative density systems [49]. Our systems with smaller ligaments, on the other hand, exhibit higher dislocation densities, possibly related due to the combined effect of more frequent defect pileups at junctions (Figure 6a, 6d, and S6) and the higher number of junctions in smaller ligament systems [44]. However, at elevated temperatures, defect migration and annihilation start to dominate over defect pileup [49,50]. Compared to the stacking fault density, the dislocation density is generally less sensitive to temperature and test geometry variations.

During compression at lower temperatures, the stacking fault density consistently increases with increasing strain, with higher relative density nanoporous systems surpassing the stacking fault densities observed in the NC systems (Figure 6b). For the lower density (<50%) np-$Al_{0.1}CoCrFeNi$ systems, the stacking fault densities for the same strain values remain below those of the NC systems for both 600 K and 1273 K (Figure 6b,d). Dislocations and stacking fault densities correlate with stress-strain curve trends, especially under tension, where the reduction of defect density aligns with decreasing stress (Figure 3b, 6e, and 6f).

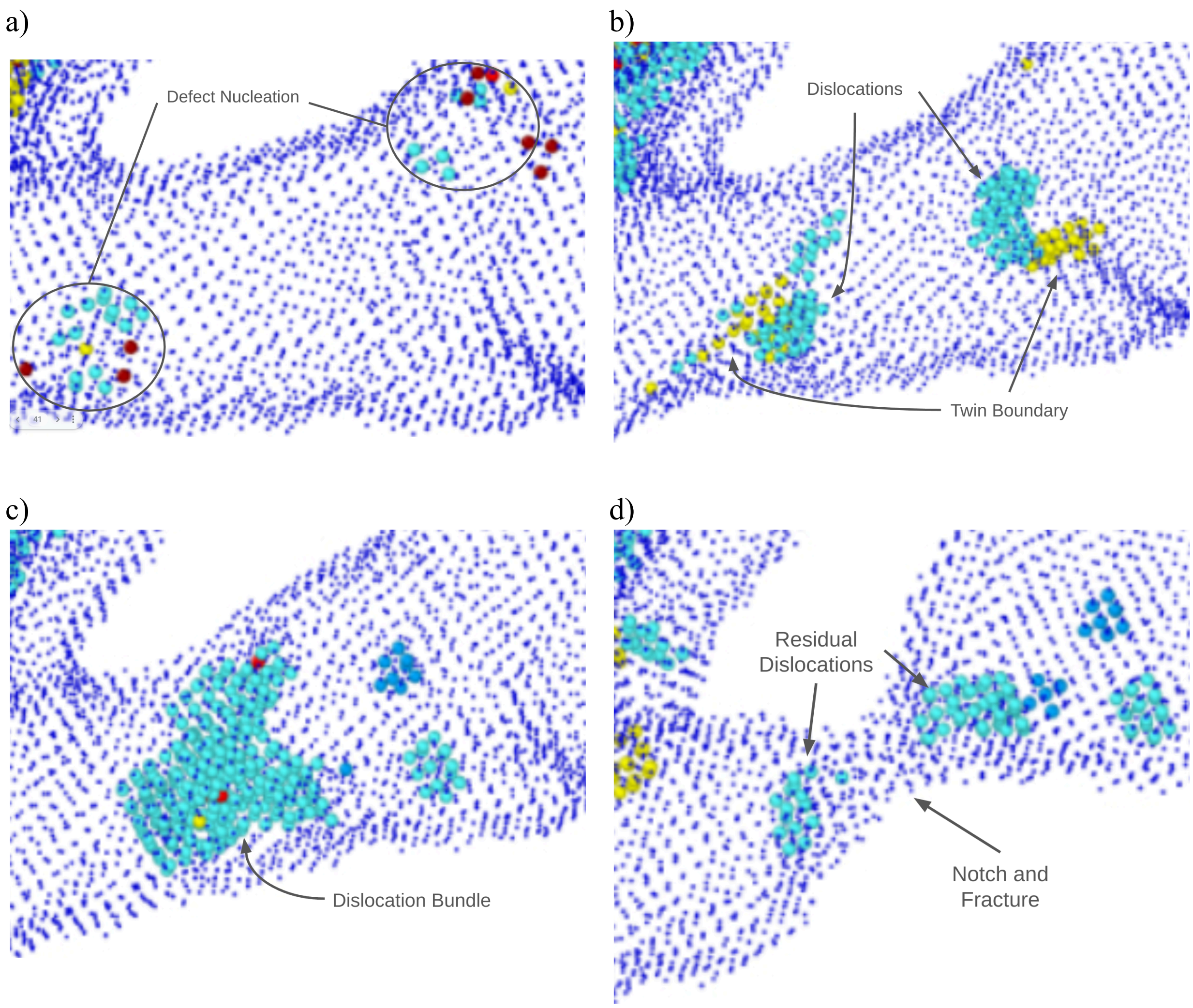

**Figure 5.** Representative fracture process of the 28.0% relative density nanoporous NbMoTaW with 3.83 nm ligaments under tensile stress. The dark blue spheres are surface atoms, light blue are vacancies, green are atoms outlining dislocations, yellow are twin boundary atoms, and red are unidentifiable atoms. The strain values are 14% (a), 15% (b), 18% (c), and 20% (d).

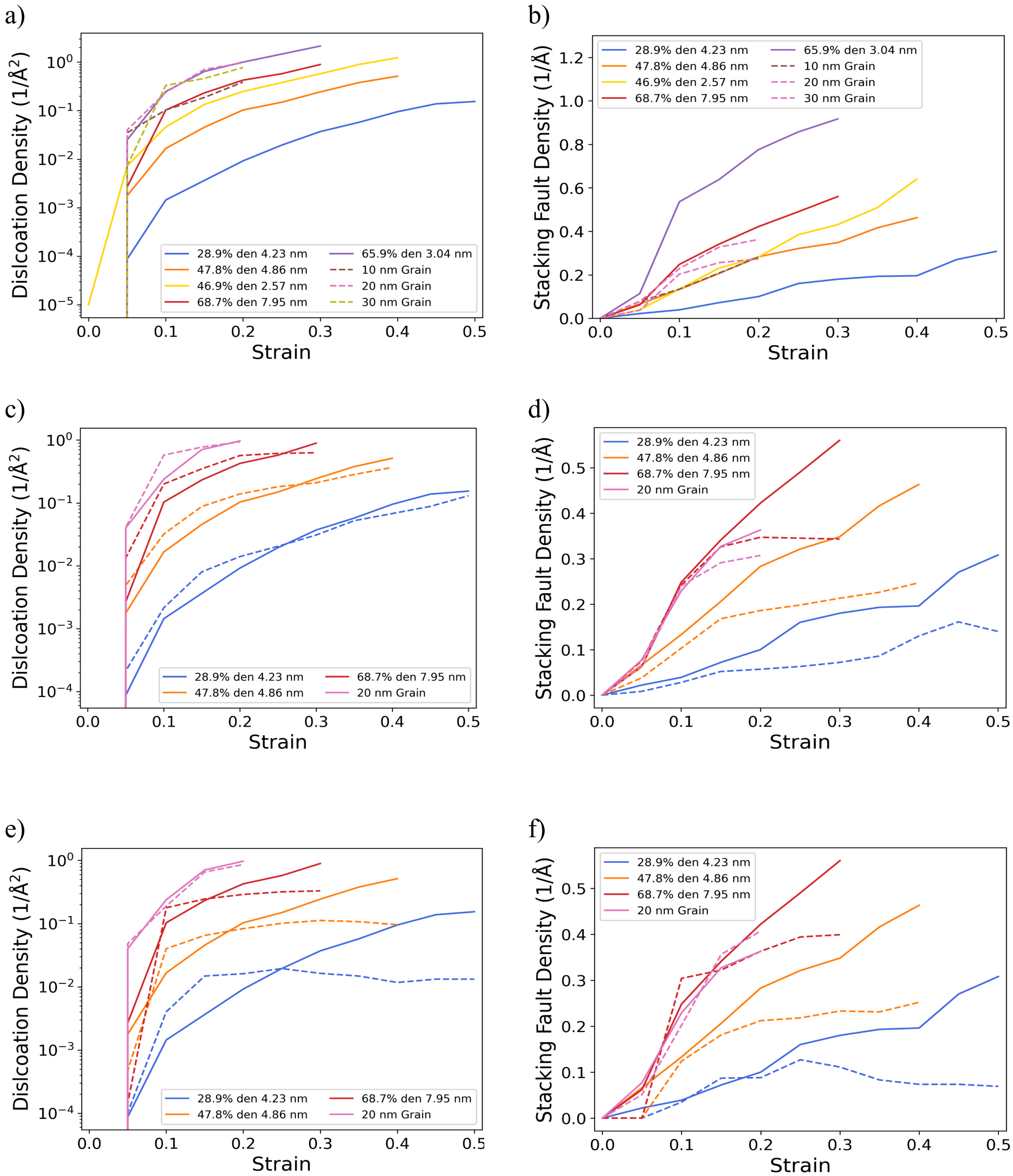

**Figure 6.** Dislocation (left) and stacking fault densities (right) vs strain for various $Al_{0.1}CoCrFeNi$ systems: defect density vs. strain for (a,b) compression of NC and NP $Al_{0.1}CoCrFeNi$ systems at 600 K; (c,d) for compression at 600 K (solid line) and 1273 K (dashed line); (e,f) compression (solid line) and tension (dashed line) at 600 K.

In contrast, none of the NbMoTaW nanoporous systems exhibited stacking fault defects. The absence of stacking faults can be attributed to the high stacking fault energy characteristic of bcc crystal structures and was found to be ~1300 mJ/$m^2$ in NbMoTaW, magnitudes lower than the $Al_{0.1}CoCrFeNi$ systems [26,51]. The dislocation density in tension is lower compared to compression (Figure S7). This disparity could be attributed to tension-compression dislocation glide velocity asymmetry, where tensile deformation could result in higher dislocation velocities, facilitating interactions and annihilation at the ligament surface [49,52]. Another explanation lies in the elongation of ligaments, which increases the surface-to-volume ratio, consequently elevating the dislocation annihilation rate. Temperature has a less pronounced effect on dislocation density in NbMoTaW compared to $Al_{0.1}CoCrFeNi$. The nanoporous NbMoTaW systems exhibit minimal changes in dislocation density with varying temperatures under tensile deformation, whereas there is a small but systematic increase in dislocation densities at higher temperatures under compression (Figure S7). The velocity of dislocations increases with higher temperature, but the formation of dislocations also becomes faster. Depending on the specific HEA, the dislocation velocity may be significantly impeded, particularly in bcc HEA materials with their slow screw dislocation motion [53]. Additionally, while screw dislocations in bcc systems are typically influenced by temperature, NbMoTaW exhibits a consistent dislocation velocity across different temperature ranges due to the presence of thermal kinks and cross kinks [54]. Under compression, it appears that the effect of higher temperatures on the rate of dislocation formation dominates over the effect of higher dislocation velocity and the associated higher annihilation rates. Further analysis of dislocation types is expounded in Supplementary Discussion 5 with observation of grain boundary induced dislocations in Figure S10.

Further investigation into the PEL of the a/2<111>{110} edge dislocation between two potential valleys in both NbMoTaW and W systems is depicted in Figure 7. The setup for these calculations is described in Supplementary Discussion 6. The analysis demonstrates that the energy barrier for dislocation motion is significantly higher in ligaments, both for NbMoTaW and W, compared to single crystal bulk configurations. This higher energy barrier for dislocation motion in ligaments may be related to the high surface-to-volume ratio in nanoscale ligaments in combination with tensile surface stress, leading to a compressive stress state in the core which makes dislocation glide more difficult [55]. Moreover, HEA systems demonstrate elevated energy barriers for both bulk and ligament structures compared to tungsten (W), offering a rationale for the remarkable ligament strength observed in nanoporous HEAs. The higher barriers in the PEL in ligaments can also be extrapolated to the calculated screw dislocations in prior investigation that demonstrates higher critical resolved shear stress compared to edge dislocation [26]. The higher energy barrier could be attributed to the entrapment of certain dislocation segments by atomic traps formed by specific local atomic configurations [56]. We conclude that the increased energy barrier in ligaments likely arises from localized stress concentrations and atomic-scale inhomogeneities, which effectively trap dislocations and impeded glide. This underscores the necessity for further exploration of the mechanisms governing dislocation motion in these materials.

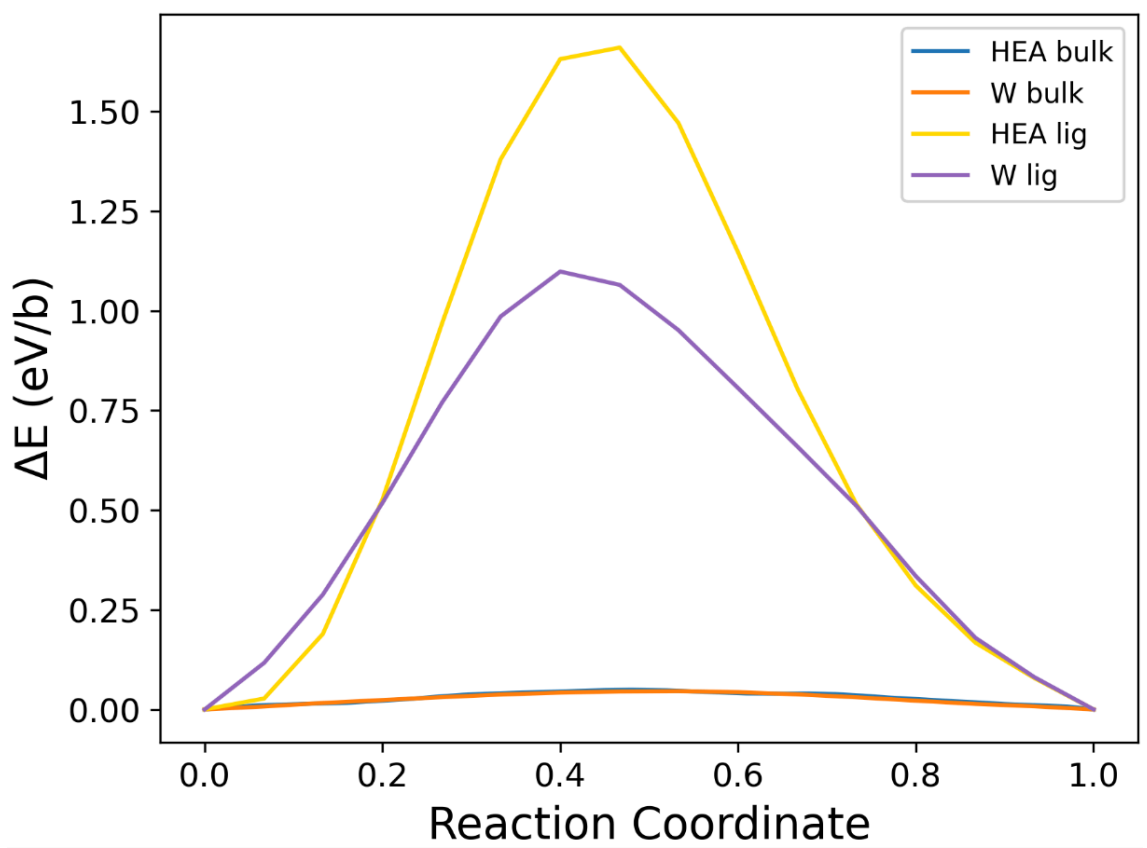

**Figure 7.** Potential energy landscape in units of eV per unit length of a/2<111>{110} edge dislocation in NbMoTaW and W. It is analyzed in both single crystal and 5 nm ligament diameter configurations.

### 3.5 Surface Energy Analysis

The surface energies (Table 6) of the nanoporous and SC $Al_{0.1}CoCrFeNi$ systems was determined through a multi-step process. Initially, we computed the total energy of the nanoporous $Al_{0.1}CoCrFeNi$ system and compared it to an equivalent bulk system with the same atom count at specified temperatures. We then divided the energy difference by the surface area of the nanoporous HEA systems, as detailed in Supplementary Table 2. Surface area calculations utilized OVITO software with a probing radius of 4 Å [45]. Surface energy for different facets of SC are calculated by taking a bulk system of $Al_{0.1}CoCrFeNi$ with the preferential [100], [110], and [111] faces in the x direction of the simulation cell and determining the total energy with and without a vacuum layer of 2 nm. The energies of the particular facets are subtracted by the bulk and divided by the surface area. The values of the bulk surface energies are within reasonable values of other DFT calculations of the individual elements of $Al_{0.1}CoCrFeNi$ [57]. Notably, smaller ligament sizes displayed approximately twice the surface energy of larger ligaments. This trend is attributed to surface energy being influenced by the curvature of ligaments, which, in turn, is associated with an increased density of step edges with greater curvature [58].

An intriguing phenomenon observed is the increase in surface energy with rising temperatures for all relative densities. This behavior contrasts with bulk metals, where the surface energy typically decreases with increasing temperature [59]. Zhang et al. identified the role of surface energy in enhancing the strength of nanoporous materials [60]. While surface energy and surface stress are distinct concepts, their interplay influences dislocation behavior near surfaces [61]. HEAs may exhibit elevated surface energies without a direct correlation to surface stress. Elevated surface energy may promote surface reconstruction or faceting, potentially altering roughness and influencing dislocation mobility. However, the exact connection requires further analysis. The potential for reduced ligament coarsening in nanoporous $Al_{0.1}CoCrFeNi$ compared to single-element nanoporous materials, which could otherwise compromise the functional properties associated with high surface energy, might partially account for this effect [62]. This effect is consistent with observations in TiVNbMoTa and TaMoNbVNi, which show reduced coarsening rates at elevated temperatures compared to expectations [62,63]. A plausible explanation is the inhomogeneous potential energy landscape in HEAs, which can hinder surface diffusion via deep atomic traps, similar to their effect on dislocation mobility [63]. Further studies are needed to fully comprehend this phenomenon. Additional investigation of surface energy as a function of temperature and mechanical testing is discussed in Supplementary Discussion 7 with accompanying surface energy vs. strain graphs in Figure S11.

## 4. Discussion

To highlight the mechanical properties of the nanoporous HEAs, a comparative analysis is conducted against other nanoporous materials (Figure 8) [5,31,64–74]. The rationale for juxtaposing MD simulations of nanoporous materials with experimental data is explained in greater detail in Supplementary Discussion 8. The yield stresses exhibited by the nanoporous HEA systems surpasses that of other studied nanoporous materials with similar relative densities. This combination of high yield strength and low-density results in an exceptionally high specific yield strength, reaching values up to five times greater than typical nanoporous materials. For further comparisons between fcc and bcc nanoporous systems, Supplementary Discussion 9 provides additional insights.

The superior performance of nanoporous HEAs may stem from a synergistic effects, possibly related to the dislocation starvation mechanism observed in nanostructured materials [75], coupled with the high strength of HEAs related to the rough PEL which reduces the dislocation mobility. The latter contribute to the strain hardening of individual ligaments, microscopically addressing the macroscopic brittleness often associated with nanoporous materials by mitigating the risk of cascading single ligament failure [53,76].

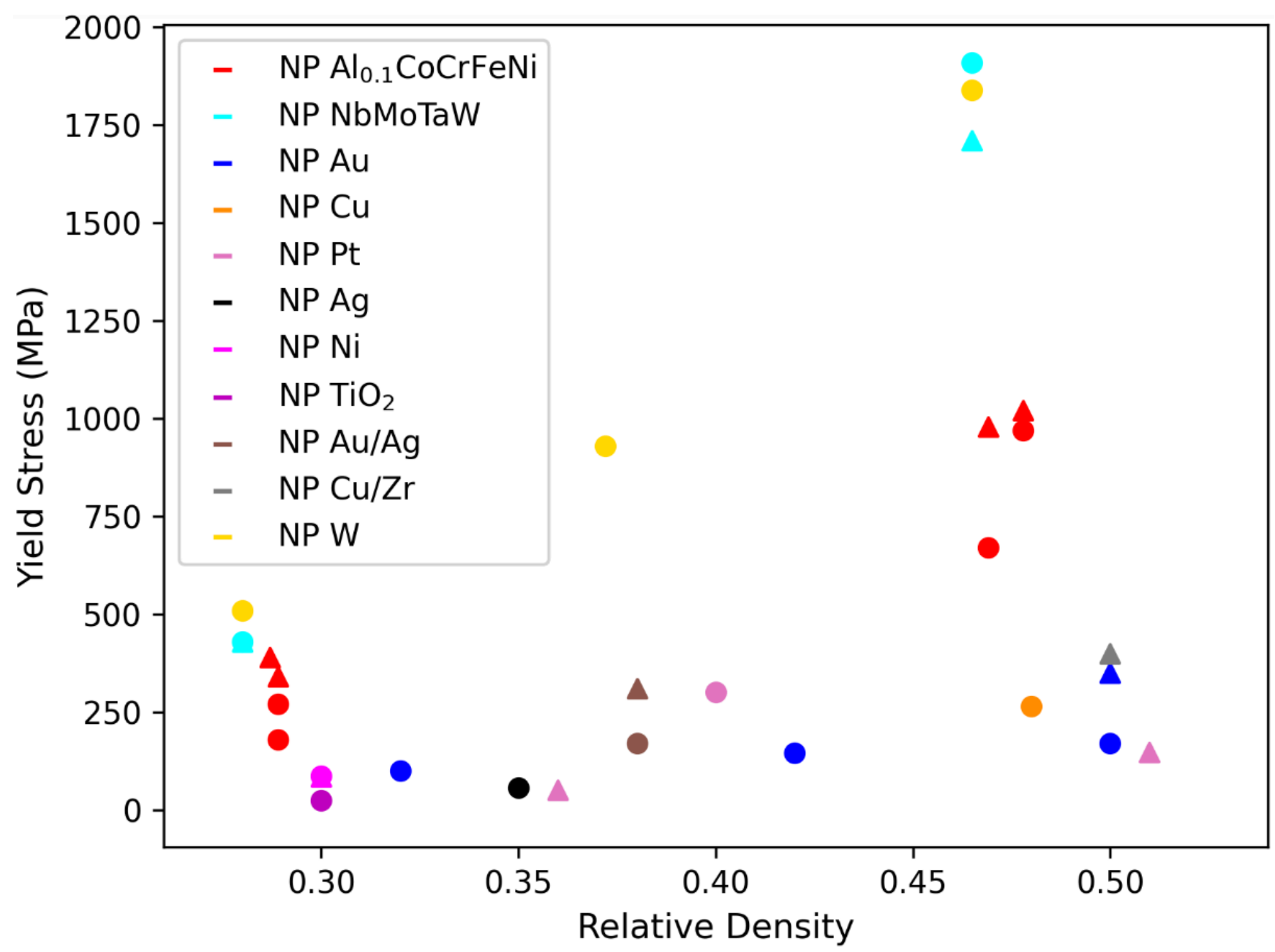

**Figure 8.** Stress vs. Relative Density of different nanoporous materials given compression (circle) or tensile (triangle) stress. These include nanoporous $Al_{0.1}CoCrFeNi$, NbMoTaW, Au [5,65–67], Cu [68], Pt [69,70], Ag [71], Ni [77], $TiO_2$ [72], Au/Ag [73], Cu/Zr [74], and W [64].

The dislocation starvation mechanism is a common phenomenon occurring in nanoporous structures that contributes significantly to their high microscopic strength by effectively annihilating dislocations at surfaces. This mechanism results in low dislocation densities, thus reducing the plastic flow within ligaments characterized by a high surface-to-volume ratio. This preservation of ligament strength becomes evident when comparing the lower dislocation densities in nanoporous structures to those in nanocrystalline systems, as illustrated in Figure 6a. The nodes are observed to have the highest defect density in the nanoporous structures (as shown in Figure S6) due to their lower surface-to-volume ratio compared to the ligaments.

Additionally, sluggish dislocation motion is another mechanism contributing to the high strength of nanoporous HEAs. This phenomenon arises from the naturally rough PEL characterizing HEAs, which features higher mean migration barriers compared to conventional alloy compositions. This rugged PEL is believed to be a primary factor behind the strain hardening effect observed in HEAs [78]. The sluggish dislocation motion influences the critical resolved shear stress (CRSS) of a/2<111>{110} edge dislocations in NbMoTaW, which is found to be approximately 320 MPa—significantly greater than the highest individual constituent value of 76 MPa in Mo [26]. This sluggish dislocation mechanism is also evident in the investigation of the potential energy landscape of edge dislocations in NbMoTaW, as depicted in Figure 7. While this particular composition exhibits sluggish diffusion for edge dislocations, other defect mechanisms, such as vacancy migration barriers, do not display such remarkable changes when compared to single-element materials. The influence of these mechanisms depends on the specific alloy composition [78,79]. This provides an opportunity to tune the mobility of defects and thus mechanical properties by engineering the PEL through the elemental compositions [55]. Compressive stress in the ligament core due to tensile surface stress may amplify the effects of the HEA's rough PEL on dislocation motion, further enhancing mechanical strength. This internal compressive stress is demonstrated in bulk and 5 nm ligaments of NbMoTaW, as depicted in supplementary Figure S8, highlighting the enhanced resistance to deformation in ligaments.

Sluggish dislocation motion in HEAs also gives rise to additional defect mechanisms that facilitate strain hardening, which varies depending on whether the crystal structure is fcc or bcc. In the case of $Al_{0.1}CoCrFeNi$ ligaments, stacking faults and partial dislocations move primarily in the <211> direction until they come in contact with a surface, where one of the partial dislocations is annihilated. After this annihilation, the stacking fault becomes trapped within the ligament due to attractive forces between the second partial dislocation and the stacking fault as can be observed in Fig. 4. This inhibition of partial dislocation-mediated stacking fault slip significantly enhances work hardening and may be a common characteristic in other fcc HEA nanoporous materials with low SFE [80,81]. This suggests a potential correlation between ligament size and average stacking fault size for fcc nanoporous HEAs. This differs from nanoporous and nanopillar gold systems, which exhibit a mix of dislocations, perfect dislocations, and perfect dislocation loops, facilitating deformation. These deformation mechanisms are accompanied by other larger defects such as twins and stacking faults [82,83]. Tables 1 and 2 indicate that the yield strength of the ligaments in nanoporous $Al_{0.1}CoCrFeNi$ increases with increasing ligament diameter, contrary to the common belief that smaller ligament sizes lead to higher strength [84]. This observation of smaller ligaments exhibiting decreased strength beyond a critical width was noted in nanopillar Titanium and is attributed to surface stress and thermal vibrations [85,86]. This suggests that the critical ligament size is between the smaller and larger ligament diameter sizes.

Regarding the NbMoTaW system, the sluggish dislocation mechanism renders screw dislocation mobility and edge dislocation mobility much more comparable than in single-element materials [26,87]. The almost complete lack of dislocations in the ligaments may be attributed to the preferential nucleation of dislocations on surfaces oriented in the {111} direction [88]. This orientation is primarily found in the nodes of the nanoporous structure where ligaments connect, whereas the surface of the ligaments typically exhibits {100} or {110} orientations. Dislocation nucleation in the nodes, coupled with comparable velocities of screw and edge dislocations, elevates the chances of dislocation pileups within the nodes. This phenomenon enhances material strength through dislocation forest hardening, consequently diminishing the likelihood of dislocations exiting the nodes to distort the ligaments and contributing to strain hardening [89].

## 5. Conclusions

Our molecular Dynamics simulations provide insights into the mechanical behavior of nanoporous HEAs, demonstrating their ability to mitigate the inherent brittleness of nanoporous materials by enhancing ligament strength and reducing the likelihood of single ligament failure cascades. These materials exhibit a combination of high specific strength, low specific modulus, and low density that carves a distinctive niche in the specific modulus-strength phase space. The simulations indicate that the elevated strength of nanoporous HEAs stems from a dual mechanism involving dislocation starvation and sluggish diffusion of dislocations. The impact of reduced dislocation mobility depends on whether the HEA possesses a face-centered cubic (fcc) or body-centered cubic (bcc) structure. In fcc structures, specifically edge dislocations are slowed down, increasing the likelihood of stacking fault formation in ligaments facilitated by the low stacking fault energy. These stacking faults then get trapped due to surface bounding and are further constrained by ligament orientation, hindering partial dislocation motion. In bcc structures, sluggish dislocation motion contributes to dislocation pileup in the nodes of the nanoporous structure, resulting in dislocation forest hardening. The comparable mobility of screw and edge dislocations fosters the interaction of numerous dislocations in the node centers, creating substantial pileups and reducing the number of dislocations traveling into ligaments where they would be needed to support deformation. The absence of dislocation nucleation in ligaments is attributed to the preferential nucleation in the {111} direction of the surfaces in the nanoporous structure, common in nodes. Additionally, the study notes that tailoring ligament size and relative density yields diverse responses in defect concentrations and surface energy. Lower relative densities are associated with reduced defect densities and higher initial surface energies. Further scrutiny of these structures promises valuable insight into the unique mechanisms at play in nanoporous HEAs, specifically engineering compositions that facilitate higher strengths and potential energy landscapes that produce high energy barriers for defect motion .

**Table 1.** Average ligament size with the respective system and relative density.

| System and Relative Density | Average Ligament Size (nm) |
|---|---|
| $Al_{0.1}CoCrFeNi$ 28.9% | 4.23 ± 1.53 |
| $Al_{0.1}CoCrFeNi$ 28.7% | 2.20 ± 0.59 |
| $Al_{0.1}CoCrFeNi$ 47.8% | 4.86 ± 1.62 |
| $Al_{0.1}CoCrFeNi$ 46.9% | 2.57 ± 0.65 |
| $Al_{0.1}CoCrFeNi$ 67.2% | 7.33 ± 1.66 |
| $Al_{0.1}CoCrFeNi$ 65.9% | 3.04 ± 0.69 |
| NbMoTaW 28.0% | 4.79 ± 1.20 |
| NbMoTaW 46.9% | 5.91 ± 1.31 |
| NbMoTaW 68.7% | 7.95 ± 1.37 |

**Table 2.** Young's modulus and yield strength of the analyzed $Al_{0.1}CoCrFeNi$ HEA systems at 298 K. Experimental and other simulation results are in parentheses. All units are in GPa.

| System | Young's Modulus Tension | Young's Modulus Compression | Yield Strength Tension | Yield Strength Compression |
|---|---|---|---|---|
| Bulk | 200.75 (199, 203 [90,91]) | 212.98 | 3.21 | 3.83 |
| 10 nm grain | 257.55 | 285.55 | 5.67 | 9.13 |
| 20 nm grain | 257.77 | 280.69 | 5.16 | 8.98 |
| 30 nm grain | 288.33 | 319.82 | 5.62 | 10.23 |
| 28.9% density 4.23 nm | 7.11 | 6.62 | 0.26 | 0.21 |
| 28.7% density 2.20 nm | 7.72 | 3.80 | 0.23 | 0.10 |
| 47.8% density 4.86 nm | 36.33 | 38.29 | 1.09 | 0.79 |
| 46.9% density 2.57 nm | 31.75 | 29.83 | 0.83 | 0.54 |
| 67.2% density 7.33 nm | 78.55 | 91.43 | 2.04 | 1.85 |
| 65.9% density 3.04 nm | 74.06 | 81.17 | 1.78 | 1.47 |

**Table 3.** Young's modulus and yield strength of the analyzed $Al_{0.1}CoCrFeNi$ HEA systems at 1273 K. All units are in GPa.

| System | Young's Modulus Tension | Young's Modulus Compression | Yield Strength Tension | Yield Strength Compression |
|---|---|---|---|---|
| Bulk | 188.13 | 191.21 | 2.03 | 2.68 |
| 10 nm grain | 221.12 | 277.11 | 4.05 | 7.83 |
| 20 nm grain | 229.07 | 240.97 | 4.12 | 7.23 |
| 30 nm grain | 251.66 | 281.38 | 4.14 | 8.44 |
| 28.9% density 4.23 nm | 3.78 | 5.20 | 0.12 | 0.06 |
| 28.7% density 2.20 nm | - | - | - | - |
| 47.8% density 4.86 nm | 24.90 | 26.32 | 0.73 | 0.40 |
| 46.9% density 2.57 nm | 10.72 | 4.76 | 0.43 | 0.10 |
| 67.2% density 7.33 nm | 62.34 | 67.14 | 1.49 | 1.34 |
| 65.9% density 3.04 nm | 55.70 | 53.22 | 1.23 | 0.79 |

**Table 4.** Young's modulus and yield strength of the analyzed NbMoTaW HEA systems at 298 K. Experimental and other simulation results are in parentheses. All units are in GPa.

| System | Young's Modulus Tension | Young's Modulus Compression | Yield Strength Tension | Yield Strength Compression |
|---|---|---|---|---|
| Bulk | 252.50 (245 [92]) | 309.82 (300 [93]) | 10.10 | 39.67 |
| 20 nm grain | 191.29 | 205.16 | 6.12 | 8.21 |
| 30 nm grain | 193.75 | 221.41 | 6.86 | 10.15 |
| 28.0% density 4.79 nm | 7.45 | 6.15 | 0.41 | 0.43 |
| 46.5% density 5.91 nm | 38.41 | 35.57 | 1.31 | 1.85 |
| 68.7% density 7.95 nm | 110.27 | 122.42 | 3.97 | 5.63 |

**Table 5.** Young's modulus and yield strength of the analyzed NbMoTaW HEA systems at 1273 K. All units are in GPa.

| System | Young's Modulus Tension | Young's Modulus Compression | Yield Strength Tension | Yield Strength Compression |
|---|---|---|---|---|
| Bulk | 190.05 | 246.99 | 7.37 | 21.24 |
| 20 nm grain | 164.14 | 177.94 | 4.94 | 6.41 |
| 30 nm grain | 170.77 | 200.97 | 6.15 | 8.40 |
| 28.0% density 4.79 nm | 5.29 | 5.23 | 0.34 | 0.36 |
| 46.5% density 5.91 nm | 30.29 | 28.18 | 1.21 | 1.47 |
| 68.7% density 7.95 nm | 90.13 | 100.17 | 3.06 | 4.41 |

**Table 6.** Surface energy of nanoporous and SC $Al_{0.1}CoCrFeNi$ systems with varying temperatures.

| System | Surface Energy 298 K ($J/m^2$) | Surface Energy 600 K ($J/m^2$) | Surface Energy 900 K ($J/m^2$) | Surface Energy 1273 K ($J/m^2$) |
|---|---|---|---|---|
| 28.9% density 4.23 nm | 1.213 | 1.227 | 1.249 | 1.243 |
| 28.7% density 2.20 nm | 2.072 | - | - | - |
| 47.8% density 4.86 nm | 1.050 | 1.072 | 1.106 | 1.126 |
| 46.9% density 2.57 nm | 1.601 | 1.638 | 1.657 | 1.729 |
| 67.2% density 7.33 nm | 0.643 | 0.757 | 0.805 | 0.851 |
| 65.9% density 3.04 nm | 1.350 | 1.417 | 1.465 | 1.508 |
| SC [100] | - | 1.748 | - | 1.807 |
| SC [110] | - | 1.924 | - | 1.945 |
| SC [111] | - | 1.543 | - | 1.551 |

Acknowledgments:

Work at LLNL was performed under the auspices of the U.S. Department of Energy by LLNL under contract No.DE-AC52-07NA27344.

This work was supported by the Nuclear Regulatory Agency [grant number 31310019M0045].

Contributions:

W. J.: Formal Analysis, Investigation, Software, Visualization, Writing-original draft, writing-review and editing.

B. J.: Writing-review and editing, Conceptualization, Project Administration

H. C.: Conceptualization, Funding Acquisition, Methodology, Project Administration, Resources, Supervision, Writing-review and editing

Competing Interest:

The authors declare no competing interest

Data Availability:

The data set used and/or analyzed during the current study are available upon reasonable request.

Supplementary Information is available for this paper.

Supplementary Materials

**Supplementary Discussion 1:**

To assess the broader impact of these materials, nanoporous High Entropy Alloy (HEA) is compared to Steel DP600, a widely used component in automobile manufacturing [1]. The nanoporous $Al_{0.1}CoCrFeNi$ system, with a relative density of 28.9% and a ligament size of 4.23 nm, has a calculated density of 2100.45 kg/m$^3$. Its specific strength and specific modulus are 0.159 MPa/(kg/m$^3$) and 0.00769 GPa/(kg/m$^3$), respectively. In contrast, Steel DP600 has a significantly higher density of 7800 kg/m$^3$, with specific strength and specific modulus values of 0.046 MPa/(kg/m$^3$) and 0.0273 GPa/(kg/m$^3$), respectively.

The nanoporous HEA not only demonstrates superior specific strength and modulus, enhancing safety performance, but also has a density that is just 26.9% of DP600's density. This significant reduction in density can contribute to lower fuel consumption and decreased $CO_2$ emissions. Studies indicate that for every 100 kg reduction in vehicle weight, $CO_2$ emissions decrease by 11 g/km [2]. Since steel constitutes approximately 70% of a vehicle's total weight, the average steel content in a car is estimated at 1319.59 kg, based on an average vehicle weight of 1885.13 kg [2,3]. Replacing steel with nanoporous HEA would result in a weight reduction of 964.24 kg, bringing the vehicle's weight down to 49% of its original mass. This translates to a reduction in carbon emissions of 106.07 g/km and a fuel consumption savings of 3.86 L per 100 km, equating to a 40% decrease in fuel usage.

Moreover, the remarkably high density of nanopores significantly increases the surface-to-volume ratio, enhancing radiation tolerance. In nanoporous HEAs, surfaces function as unsaturable sinks for defects, enabling a self-healing mechanism that makes them highly promising for next-generation nuclear reactors. Notably, materials such as NbMoTaW, when used for additional shielding and as fusion first-wall components, demonstrate remarkable helium radiation resistance [4-7].

**Supplementary Discussion 2:**

The phase separation was achieved through demixing of the fcc Ni-Cu system, where atomic concentrations were determined based on the relative density. This process utilized a lattice parameter of 0.356 nm and an embedded atom method (EAM) interatomic potential for Ni-Cu [8,9]. Additionally, a fully repulsive Lennard-Jones potential was implemented to model atomic interactions between Ni and Cu, ensuring complete immiscibility within the system [10]. A Nose/Hoover thermostat ensemble was applied at a temperature exceeding the Ni-Cu system's melting point for a predefined duration to induce phase separation. Extended demixing times led to increased ligament sizes, with a 100 ps demixing period used for larger ligaments and a 50 ps duration for smaller ligament structures. Upon completion of the process, the Cu phase was removed, leaving behind a Ni scaffold, which served as the framework for randomly placing atoms comprising the disordered fcc crystalline phase with an $Al_{0.1}CoCrFeNi$ composition.

The structures shown in Figures S6 and S9 do not exhibit true bimodal porosity. The fine internal texture is due to atomic-scale roughness that arises from the non-equilibrium dealloying-like simulation process. These features reflect local curvature and surface faceting, not secondary pores. We estimate this surface roughness increases the specific surface area by ~15-25% compared to an ideal bicontinuous structure with smooth ligaments. Simulation cell dimensions and resolution details are provided in Supplementary Discussion 5.

A similar approach was employed using a W-Ta binary system with a lattice parameter of 0.316 nm to generate the nanoporous NbMoTaW system [11]. While this method may not directly correspond to a physically realistic means of experimentally producing nanoporous materials, it provides a suitable bicontinuous structure necessary for simulation analysis.

**Supplementary Discussion 3:**

The energy barrier calculations for the $a/2\langle 110\rangle\{111\}$ edge dislocation in the $Al_{0.1}CoCrFeNi$, Ni, and Co systems were conducted using the nudged elastic band method. In this approach, a slab is oriented with x//$[1\bar{1}0]$, y//[111], and z//$[11\bar{2}]$ directions for the bulk material, with dimensions 12.50nm x 12.37nm x 13.12nm.

Atomsk software is used to introduce an edge dislocation at the center of the simulation box by removing a half-plane below the glide plane while maintaining periodicity along the x-axis [20]. A second identical system is created, but with an edge dislocation shifted by one y lattice parameter unit. Both systems undergo minimization with periodic boundary conditions applied in the x and z directions and shrink-wrapped boundary conditions in the y direction. After minimization, the nudged elastic band method is applied to determine the potential energy landscape between the two reaction coordinates, expressed in eV per unit length along the x-direction.

The Potential Energy Landscape (PEL) of the edge dislocations are displayed in Figure S3 where the $Al_{0.1}CoCrFeNi$ dislocation is shown to have the highest PEL compared to the single constituent materials.

**Supplementary Discussion 4:**

While the Dislocation Extraction Algorithm (DXA) is generally effective in identifying defects in fcc systems, it may be overly simplistic for bcc structures (as discussed in Supplementary Discussion 2), where it could potentially overlook defects such as planar faults and less common dislocations beyond screw and edge dislocations. To address these limitations, the bcc defect analysis (BDA) tool is utilized, leveraging the common neighbor analysis method integrated into OVITO [12,13].

Using the BDA tool, we identified minute twin boundaries within the nodes of the nanoporous system that were not detected in the initial DXA analysis. Additionally, the BDA revealed a greater number of dislocations at higher strain values compared to DXA, particularly in regions where ligament fracture was occurring as showcased in Figure S9.

**Supplementary Discussion 5:**

An analysis of the Burgers vectors in the $Al_{0.1}CoCrFeNi$ system reveals that Shockley partial dislocations with a Burgers vector of a/6<112> dominate at low strain levels in both high- and low-temperature experiments. While these dislocations remain the most prevalent, other types, including a/3<1-00> stair-rod dislocations and a/3<110> Lomer dislocations, are also observed. These dislocation types are frequently reported in experimental studies of bulk $Al_{0.1}CoCrFeNi$ [14,15].

With increasing strain, noticeable changes in dislocation behavior emerge. Ligament systems experiencing approximately 35% strain or higher begin to exhibit dislocations with Burgers vectors a/18<721> and a/18<552>, which appear to be associated with grain boundaries in the fcc lattice structure [16]. As strain continues to rise, these dislocation types progressively surpass the Shockley partials. This abrupt shift in dislocation characteristics is likely due to the formation of grain boundaries as ligaments come into contact under high compressive strain (Figure S10).

The strain threshold at which these dislocations appear is influenced by the relative density of the $Al_{0.1}CoCrFeNi$ nanoporous system, with higher relative density samples exhibiting these dislocations at lower strain values. Moreover, smaller ligament systems appear to impact dislocation development by increasing the strain required for the formation of a/18<721> and a/18<552> dislocations at comparable relative densities. This behavior mirrors observations of grain boundary formation during thermal coarsening in nanoporous Au [17].

In the NbMoTaW system, the most prevalent dislocation types observed during both tension and compression are a/2<111> edge and screw dislocations. Comparisons with other molecular dynamics and density functional theory calculations indicate the edge dislocations play a critical role in governing the strength of refractory HEAs [18,19]. At higher strain levels, a/2<111> dislocations remain dominant; however, other less common dislocation types, such as a<100> and a<110>, increase in density as strain values rise. The presence of a<110> dislocations is another characteristic feature of NbMoTaW that contributes to its strength control [19].

**Supplementary Discussion 6**:

The energy barrier calculations for the a/2<111>{110} edge dislocation in both the NbMoTaW and W systems were conducted using the nudged elastic band method. In this approach, a slab is oriented with x//$[1\bar{1}0]$, y//$[111]$, and z//$[11\bar{2}]$ directions for the bulk material, with dimensions 13.79nm x 16.75nm x 12.74nm. A nanopillar with a 5 nm diameter is then generated along the y-axis with dimensions 22.91nm x 13.89nm x 15.87nm.

Atomsk software is used to introduce an edge dislocation at the center of the simulation box by removing a half-plane below the glide plane while maintaining periodicity along the y-axis [20]. A second identical system is created, but with an edge dislocation shifted by one y lattice parameter unit. Both systems undergo minimization with periodic boundary conditions applied in the y and x directions and shrink-wrapped boundary conditions in the z direction. After minimization, the nudged elastic band method is applied to determine the potential energy landscape between the two reaction coordinates, expressed in eV per unit length along the x-direction.

**Supplementary Discussion 7:**

The surface energy appears to vary depending on the type of mechanical test performed. As shown in Figure S11, the surface energy under both compressive and tensile tests for the 600 K and 1273 K systems follows distinct patterns. In tensile tests, surface energy initially increases before stabilizing for most of the strain range, with a slight decline at higher strain levels. However, higher temperatures tend to suppress these surface energy increases, and in some cases, surface energy even decreases at higher strain levels.

In compression tests, surface energy follows a similar trend: an initial increase, followed by a plateau, and then a growth phase at higher strains. These surface energy trends under compression are consistent with previous nanoporous W simulations, which exhibit similar surface energy versus strain behavior [21].

Temperature influences surface energy differently in compression and tensile tests. At higher strain levels, increasing temperature reduces surface energy in tensile tests but enhances it in compression. While previous studies suggest that decreasing surface energy may drive a transition from twinning to perfect dislocation plasticity, no such shift in the deformation mechanism was observed for the second-highest surface energy system, which has a relative density of 46.9% and ligament size of 3.04 nm. It is likely that even higher surface energy levels would be necessary to induce such a transformation in the deformation mechanism.

It is important to highlight that the behavior of the 46.9% relative density system with a 2.57 nm ligament size deviates from other systems at higher temperatures and strain levels (Figure S11). This difference is particularly noticeable in tensile tests, where the surface energy of the 46.9% density, 2.57 nm ligament system continues to rise, whereas the other systems either plateau or decline. A similar trend is observed in compression tests, where the 46.9% density system exhibits lower surface energy at higher strains compared to the 65.9% relative density system with a 3.04 nm ligament size. These variations may be attributed to the small ligament sizes breaking down in a manner similar to the 27.8% density, 2.20 nm ligament system, which became unstable at elevated temperatures. This instability may be further intensified by the high strain conditions.

Surface energy and surface stress, while related, are distinct physical quantities. Surface energy refers to the thermodynamic cost of creating new surface area (bond breaking), whereas surface stress relates to the mechanical strain imposed on surface atoms that are not in equilibrium positions. While increased surface stress can influence roughness and defect interactions, the two are not strictly proportional. Additionally, mechanical loading can modify surface atom configurations, potentially reducing surface stress. Environmental effects (e.g., oxidation) may also alter surface stress and surface chemistry [22,23].

**Supplementary Discussion 8**:

A crucial consideration when comparing MD results with experimental findings is the significantly high strain rate inherent in MD simulations, which generally leads to higher yield strengths. This discrepancy is particularly evident when contrasting stress-strain calculations for bulk gold between experiments and simulations. In MD simulations, the yield strength is approximately 8.21 GPa, whereas experimental measurements report yield strengths around 205 MPa [24,25].

This difference is especially pronounced in bulk and nanocrystalline/polycrystalline systems, where high strain rates induce unique defect mechanisms that contribute to increased strength, such as the formation of low dislocation densities. The primary factors responsible for the elevated strength observed in MD simulations appear to be the high strain rate and the relatively small system volume [26].

However, this concern becomes less significant when studying nanoporous materials with bicontinuous structures. MD simulations of nanoporous gold and platinum show remarkable agreement with experimental data, even when accounting for high strain rates, as illustrated in the Gibson-Ashby plot in Figure 1 of the main text [27-32]. This consistency can be attributed to the fact that the ligaments in nanoporous structures behave similarly to nanowires or nanopillars, typically measuring in the tens of nanometers. As a result, the mechanical response observed in simulations closely aligns with experimental findings. Theoretical yield strengths obtained from MD simulations match well with expected experimental values, which are primarily influenced by nanopillar orientation and size [33].

The primary difference between simulation and experimental results lies in the strain value at which the yield strength is reached, with experimental values generally being lower than those observed in MD simulations [33]. Given that the MD simulations in this study involve ligaments smaller than 10 nm and employ large-scale models with over 2,000,000 atoms, it is reasonable to anticipate that experimental data under similar conditions would yield comparable results.

**Supplementary Discussion 9**:

Many of the nanoporous systems examined in this study exhibit an fcc crystal structure. The exceptions are nanoporous NbMoTaW and W, both of which demonstrate significantly higher yield strengths compared to other fcc nanoporous materials. Unlike fcc systems that deform primarily along {111} close-packed planes, bcc metals lack true close-packed planes. Instead, plasticity occurs through complex slip involving {110}, {112}, or {123} planes, often dominated by screw dislocations with non-planar cores [34].

In fcc metals, slip is governed by dislocation glide along the close-packed {111} planes in the a/2<110> directions, which offer the lowest energy pathways for plastic deformation [34]. Low stacking fault energies facilitate the splitting of dislocations into partial dislocations, which then dissociate onto these closely packed planes, making slip relatively easy compared to bcc structures [34].

The primary dislocation type in bcc structures is the screw dislocation, which inherently lacks well-defined slip planes. This constraint reduces the mobility of screw dislocations relative to faster-moving edge dislocations, ultimately increasing strength while decreasing the ductility of bcc materials [35]. By integrating a nanoporous bicontinuous architecture with a bcc crystal structure, the intrinsic brittleness of bcc materials can be mitigated by lowering the Young's modulus, following Gibson-Ashby scaling laws [36].

This concept of bcc materials exhibiting higher strengths extends to nanoporous HEA systems, which may demonstrate even greater strain hardening compared to single-element materials. In bcc HEAs, solid solution strengthening suggests that at low stress levels, common screw and edge dislocations remain immobile until a critical stress threshold is reached. This results in exceptionally high strength at low strain levels. Once this critical stress is exceeded, both dislocation types exhibit similar behavior with relatively low velocities. This contrasts with single-element bcc materials, where screw dislocations primarily govern ductility, while edge dislocations move more rapidly and contribute significantly to plastic deformation [34].

The reduced mobility of dislocations increases the likelihood of dislocation-dislocation interactions, potentially promoting dislocation multiplication and enhancing strain hardening [35]. The response to high strain varies depending on the specific bcc HEA being examined. For example, the bcc HEA $Co_{16.67}Fe_{36.67}Ni_{16.67}Ti_{30}$ exhibits a transition from jerky motion to smooth gliding dislocations at high strain levels, similar to other single-element bcc materials [35]. In contrast, the TiZrNbHfTa bcc system exhibits significant cross-slipping during fracture analysis, a behavior absent in the previously mentioned bcc HEA, thereby enhancing ductility [27]. This highlights the importance of selecting an appropriate bcc HEA to mitigate slip mechanisms that could accelerate ligament failure in a nanoporous system.

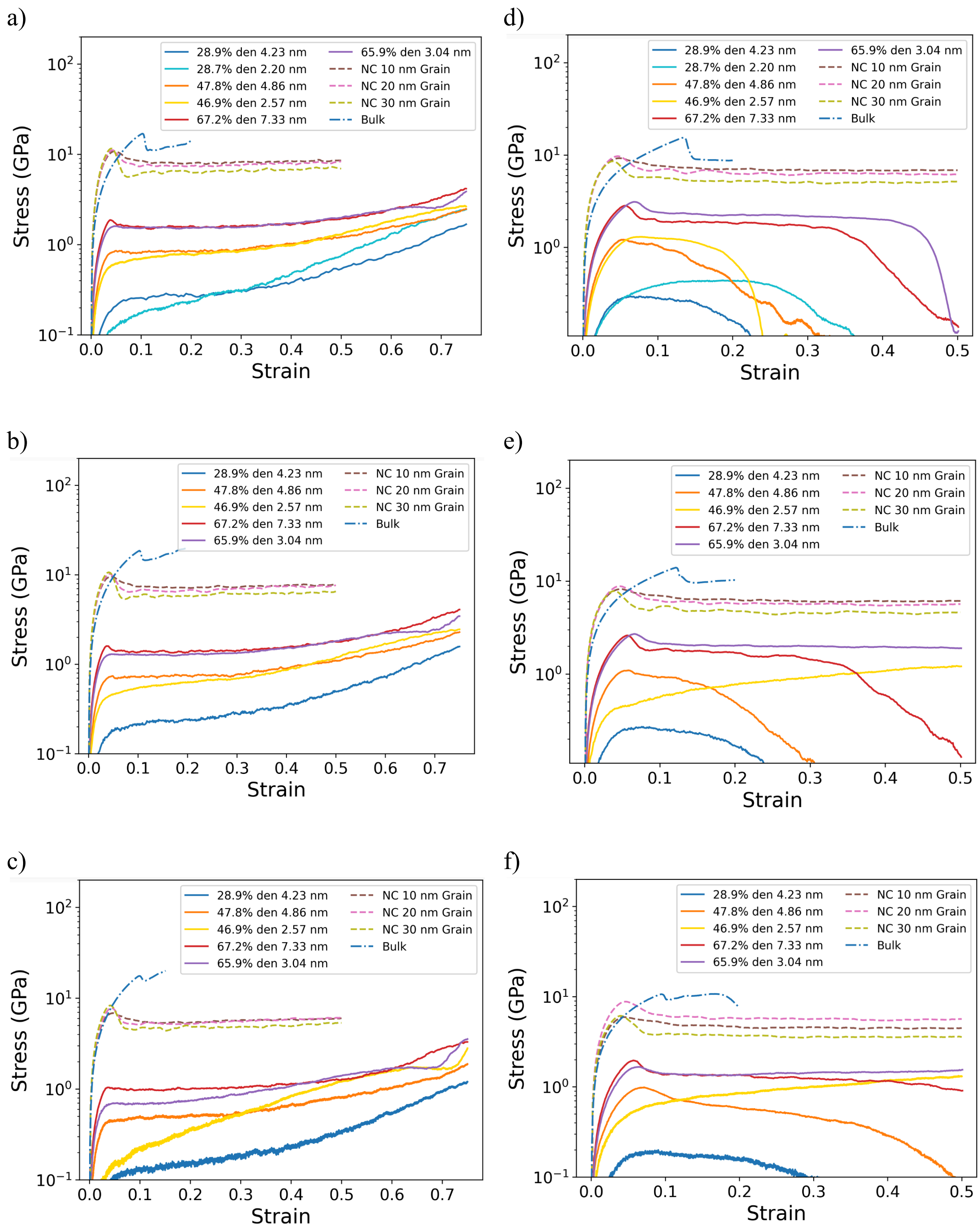

**Figure S1**. Stress strain curves of $Al_{0.1}CoCrFeNi$ systems under compression at (a) 298 K, (b) 600 K, (c) 1273 K and tension at (d) 298 K, (e) 600 K, (f) 1273 K. NC represent nanocrystalline systems.

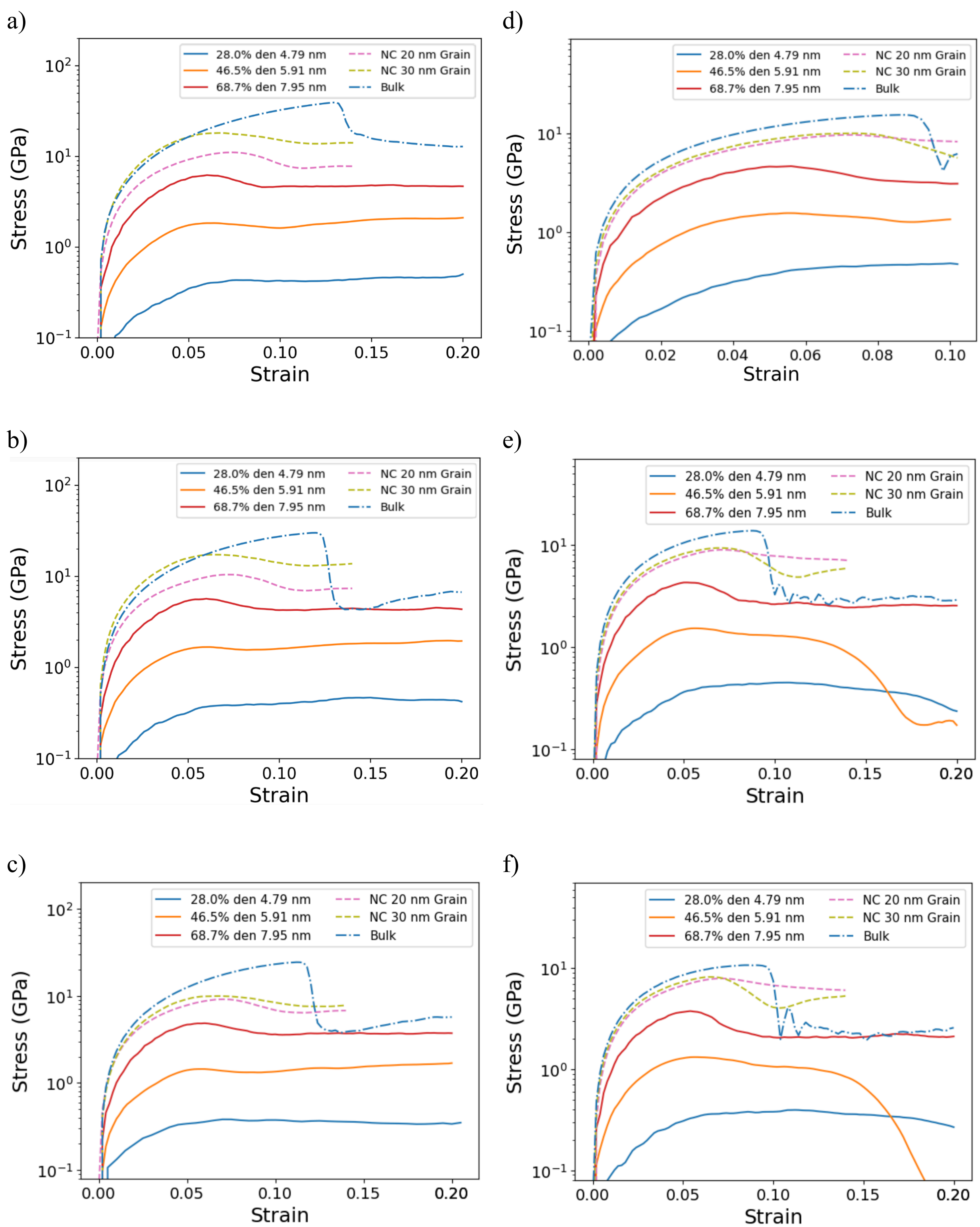

**Figure S2.** Stress strain curves of NbMoTaW systems under compression at (a) 298 K, (b) 600 K, (c) 1273 K and tension at (d) 298 K, (e) 600 K, (f) 1273 K. NC represents nanocrystalline systems.

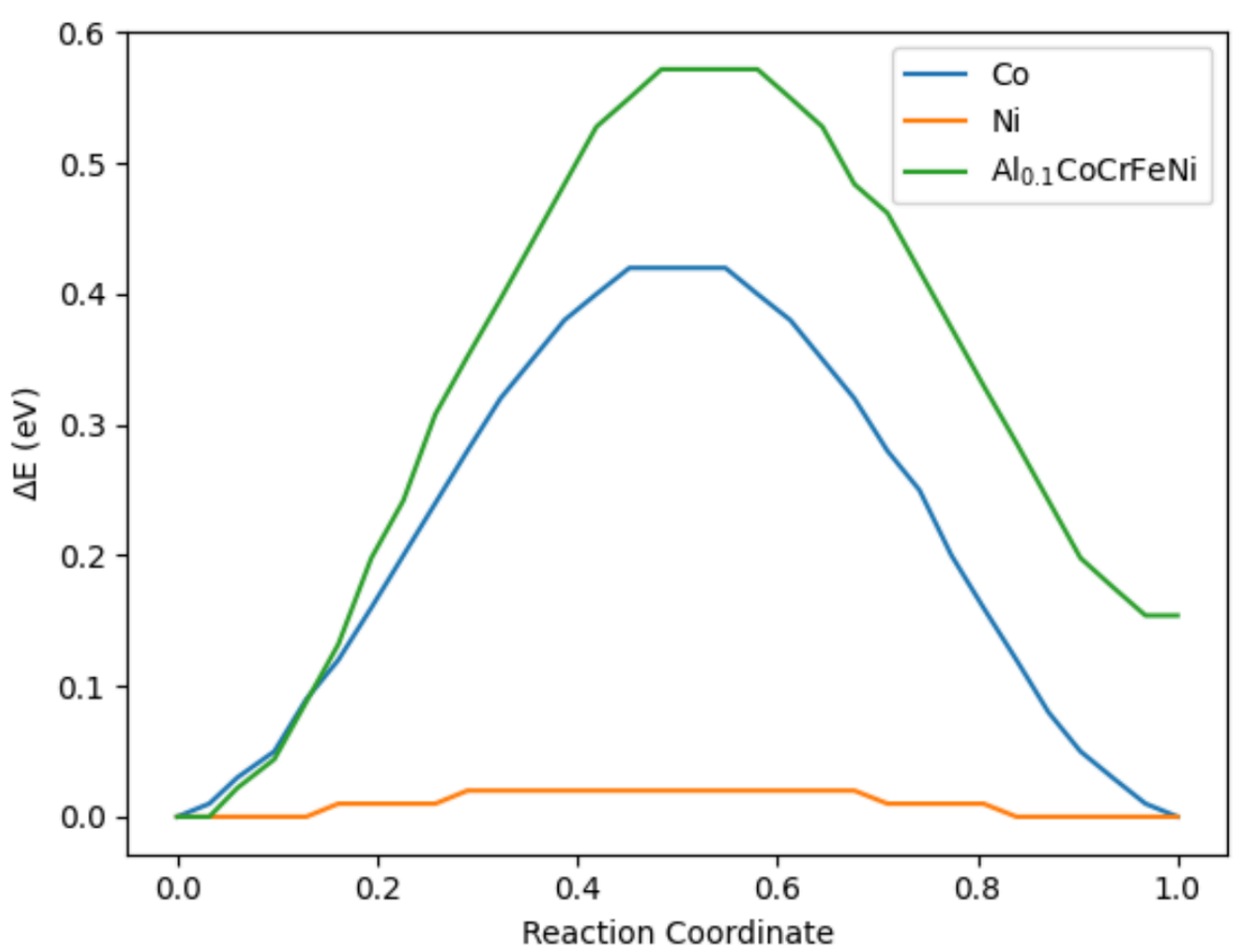

**Figure S3.** Potential energy landscape of $Al_{0.1}CoCrFeNi$, Co, and Ni in units of eV for the a/2<110>{111} edge dislocations.

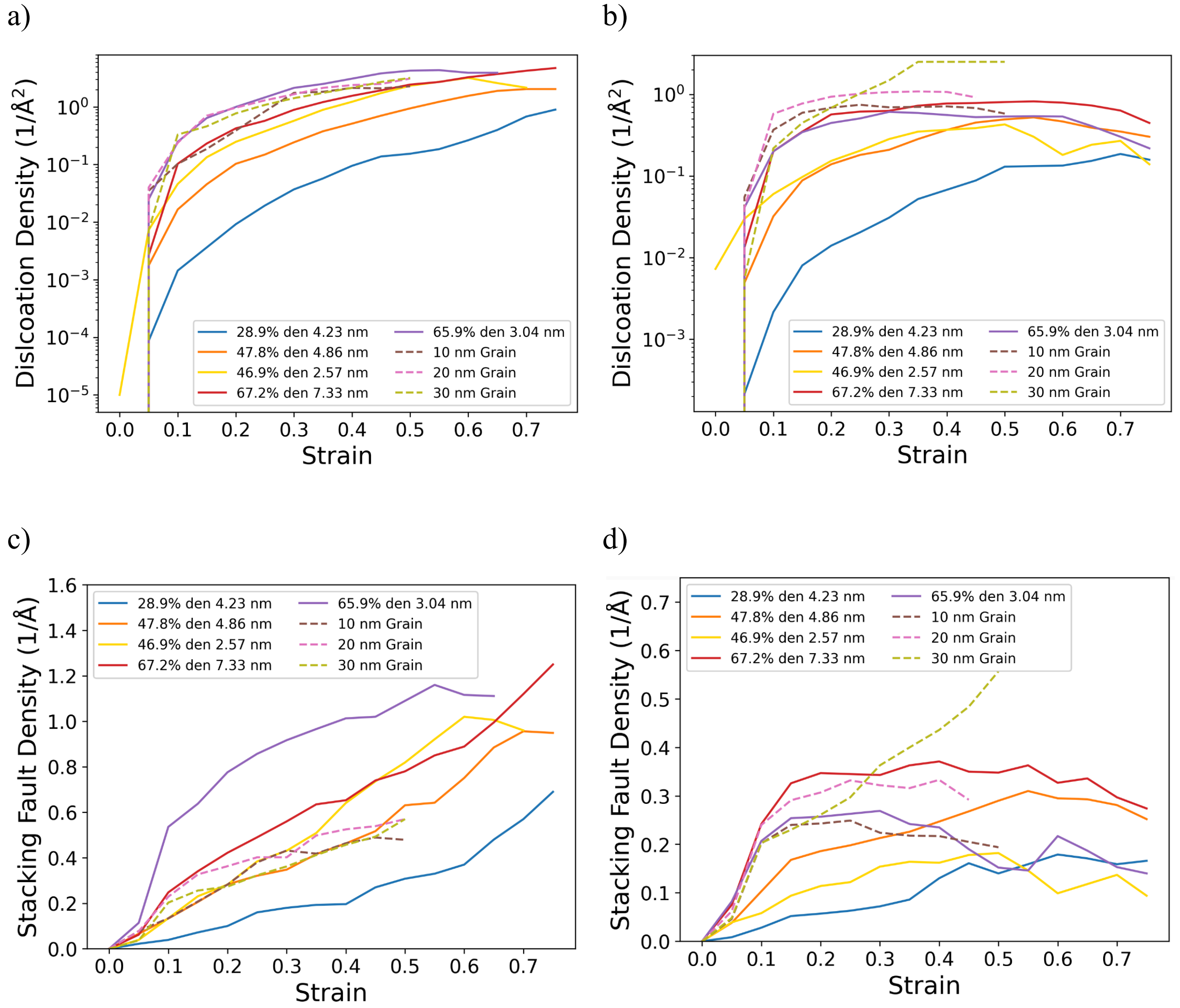

**Figure S4.** Dislocation concentration vs strain at 600 K (a) / 1273 K (b) and stacking fault concentration vs. strain at 600 K (c) / 1273 K (d) for the $Al_{0.1}CoCrFeNi$ HEA systems under compression. Nanocrystalline systems are shown as dashed curves.

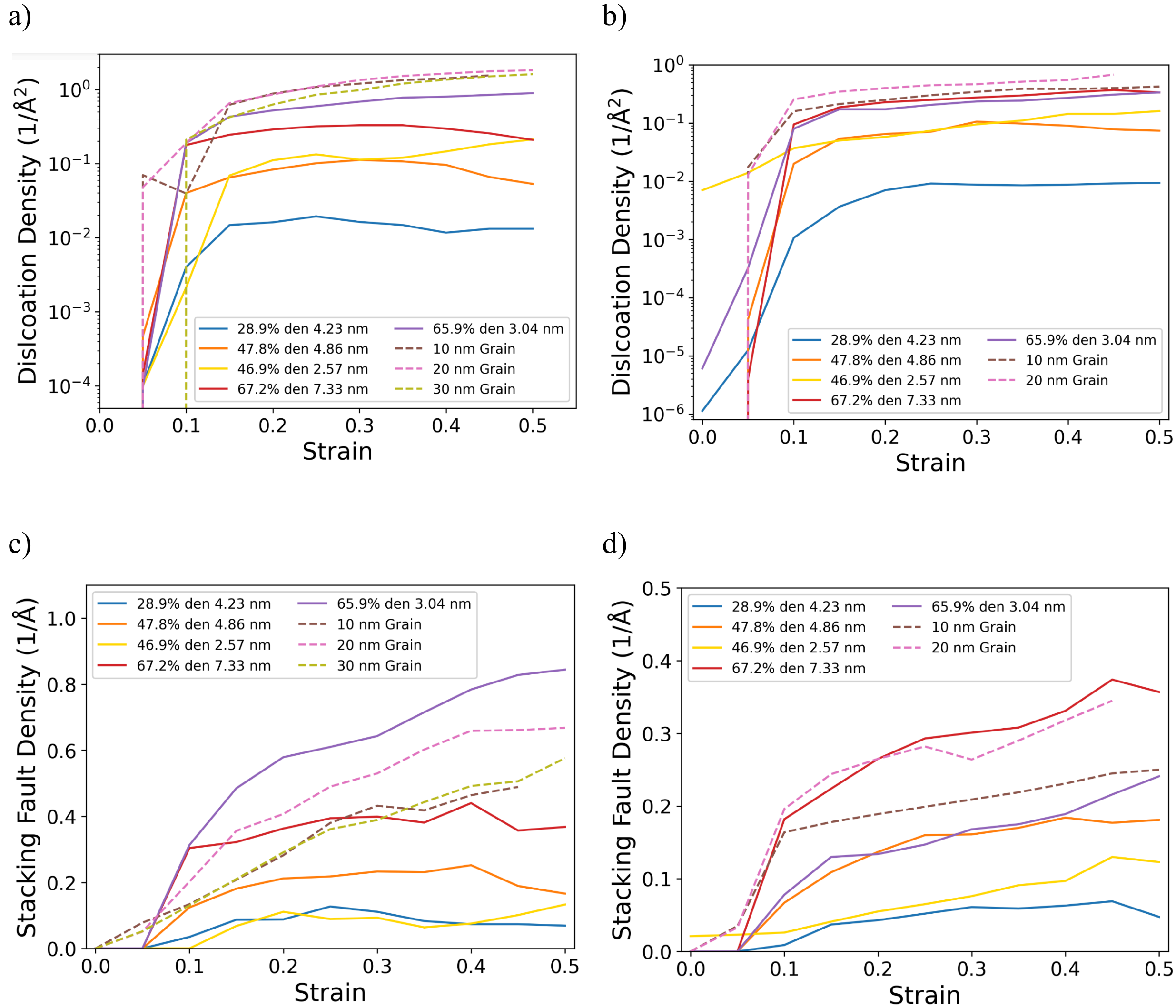

**Figure S5.** Dislocation concentration vs strain at 600 K (a) / 1273 K (b) and stacking fault concentration vs. strain at 600 K (c) / 1273 K (d) for the $Al_{0.1}CoCrFeNi$ HEA systems under tension. Nanocrystalline systems are shown as dashed curves. Note the different scale of the 1273 K plots.

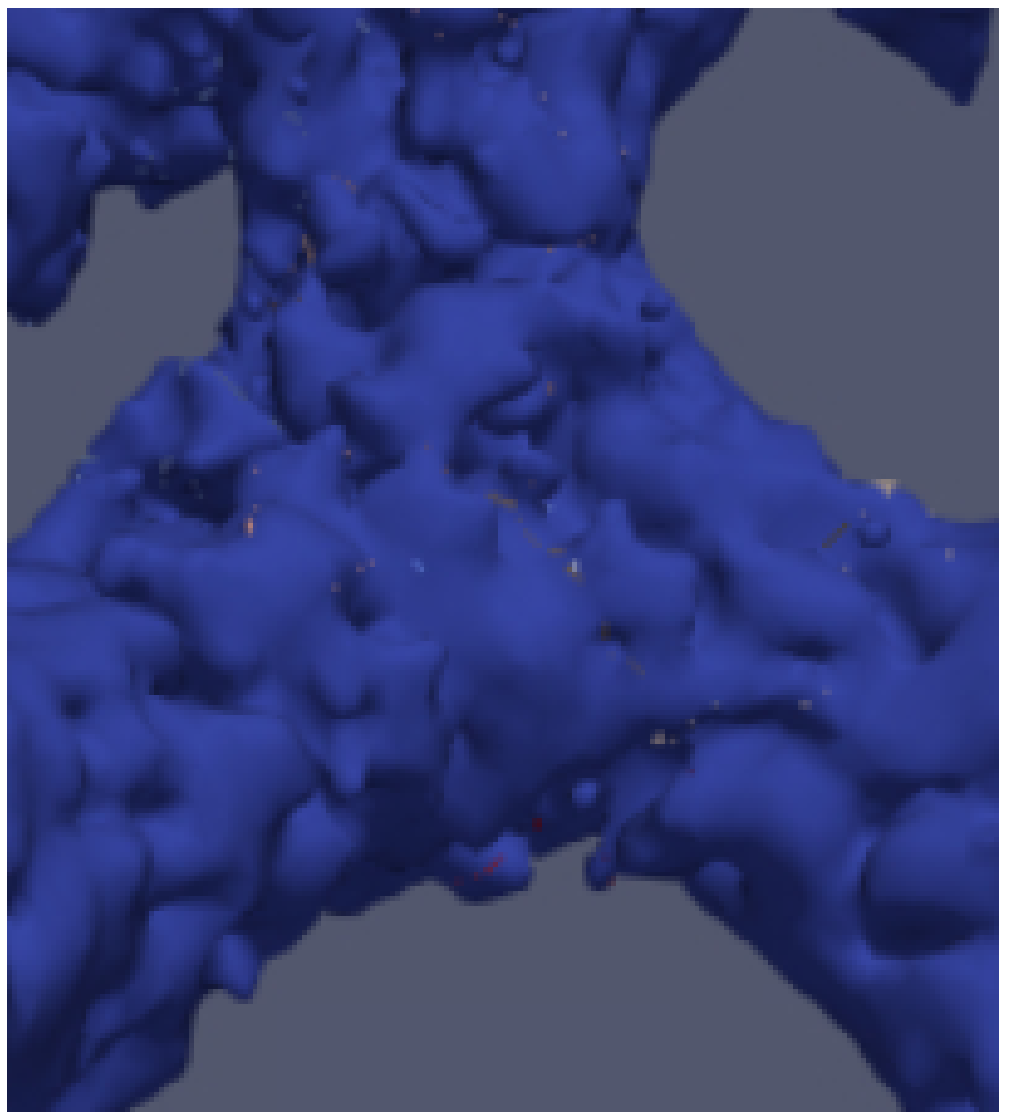

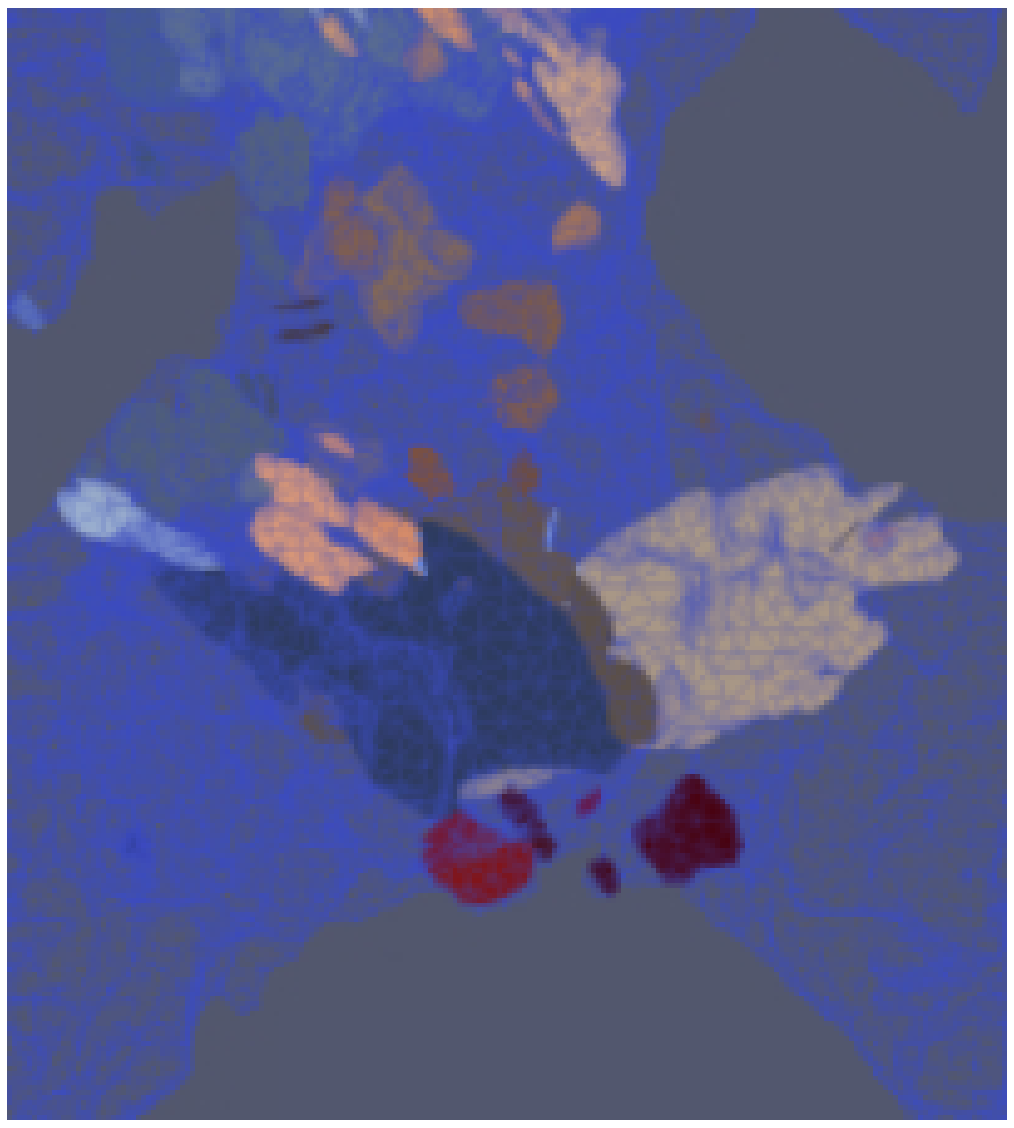

**Figure S6**. Display of defect pileup in the joints of the NP $Al_{0.1}$CoCrFeNi. The image on the left shows the surface of the NP system and the image on the right displays the defects inside the ligament that shows twin boundaries and stacking faults with different colored planes. This is the 28.9% relative density 4.23 nm ligament size nanoporous $Al_{0.1}$CoCrFeNi under tensile stress at 1273 K.

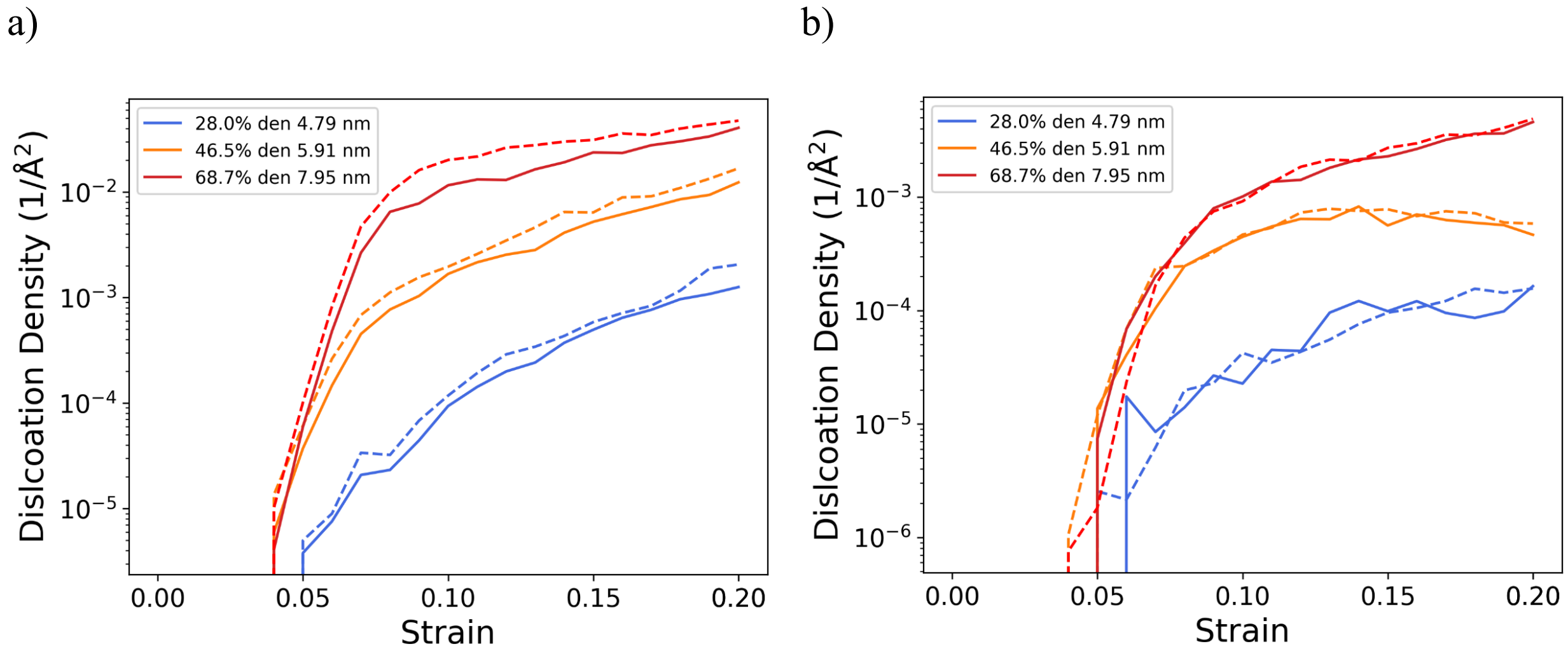

**Figure S7**. Defect Concentration vs. Strain curves for the NbMoTaW nanoporous systems that focus on dislocations at 600 K (solid line) and 1273 K (dashed line) for compression (a) and tension (b).

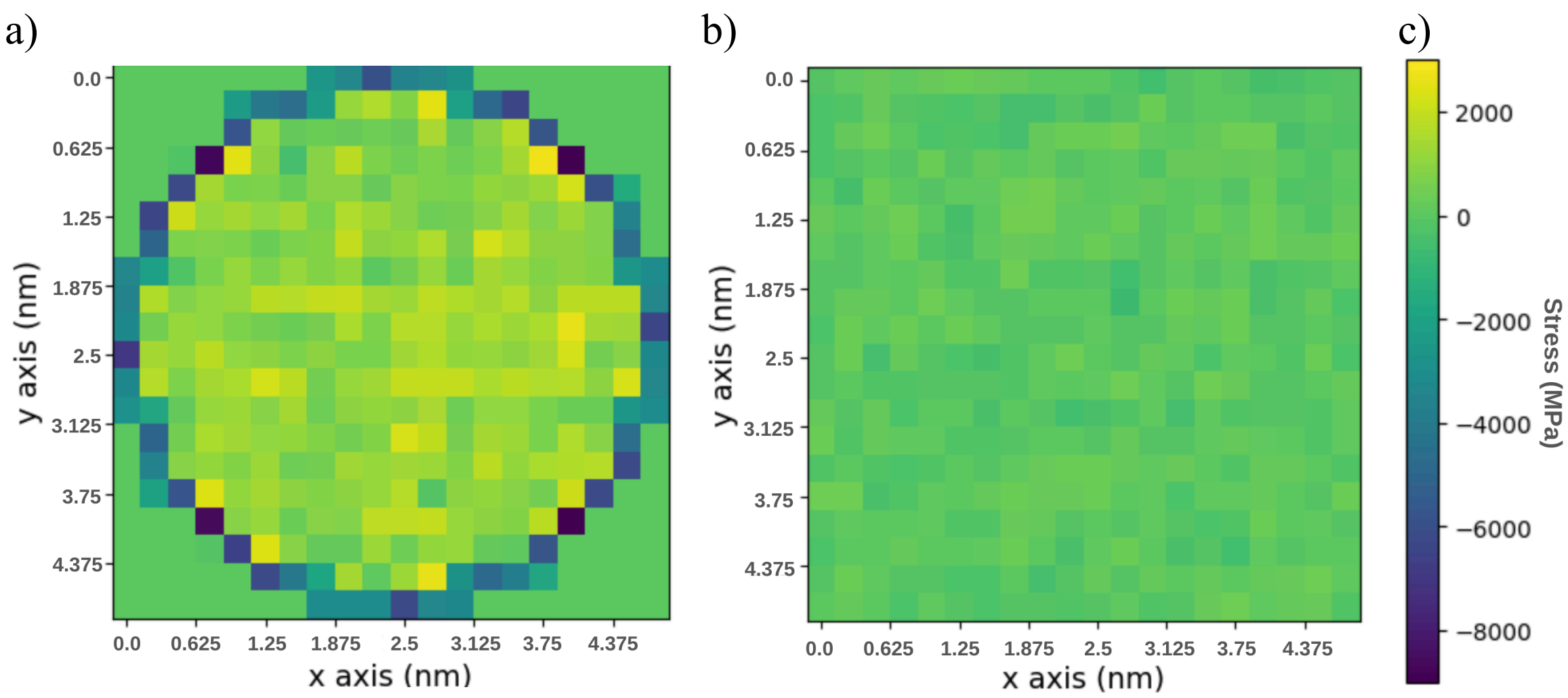

**Figure S8.** Internal stress of 5 nm ligament (a) and bulk (b) NbMoTaW with the heat map legend (c). 400 bins in the x and y direction are used to calculate the averaged stress in the z direction.

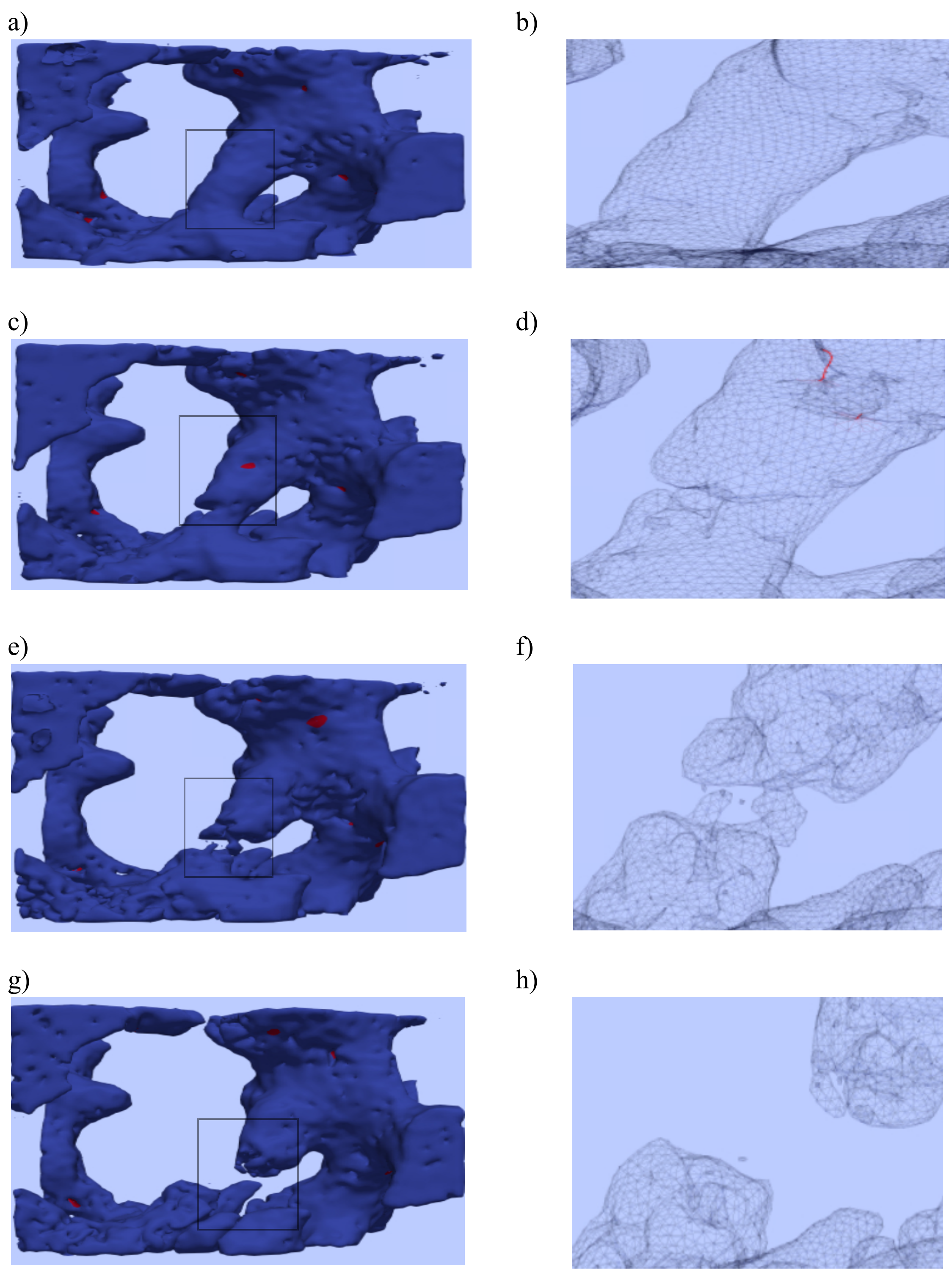

**Figure S9**. The fracture process of the 28.0% relative density 4.79 nm ligament size nanoporous NbMoTaW HEA under tensile testing. This shows the surface solid mesh outlining the atomic surface and dislocations colored red in the ligament. The strain values represented are (a,b) 13.00%, (c,d) 15.00%, (e,f) 18.00%, and (g,h) 20.00%, respectively. Note: Apparent internal

textures in the ligaments result from surface roughness rather than a second level of porosity. The structures reflect non-equilibrium morphologies and higher curvature typical of early-stage dealloying. No intentional bimodal pore distribution is present.

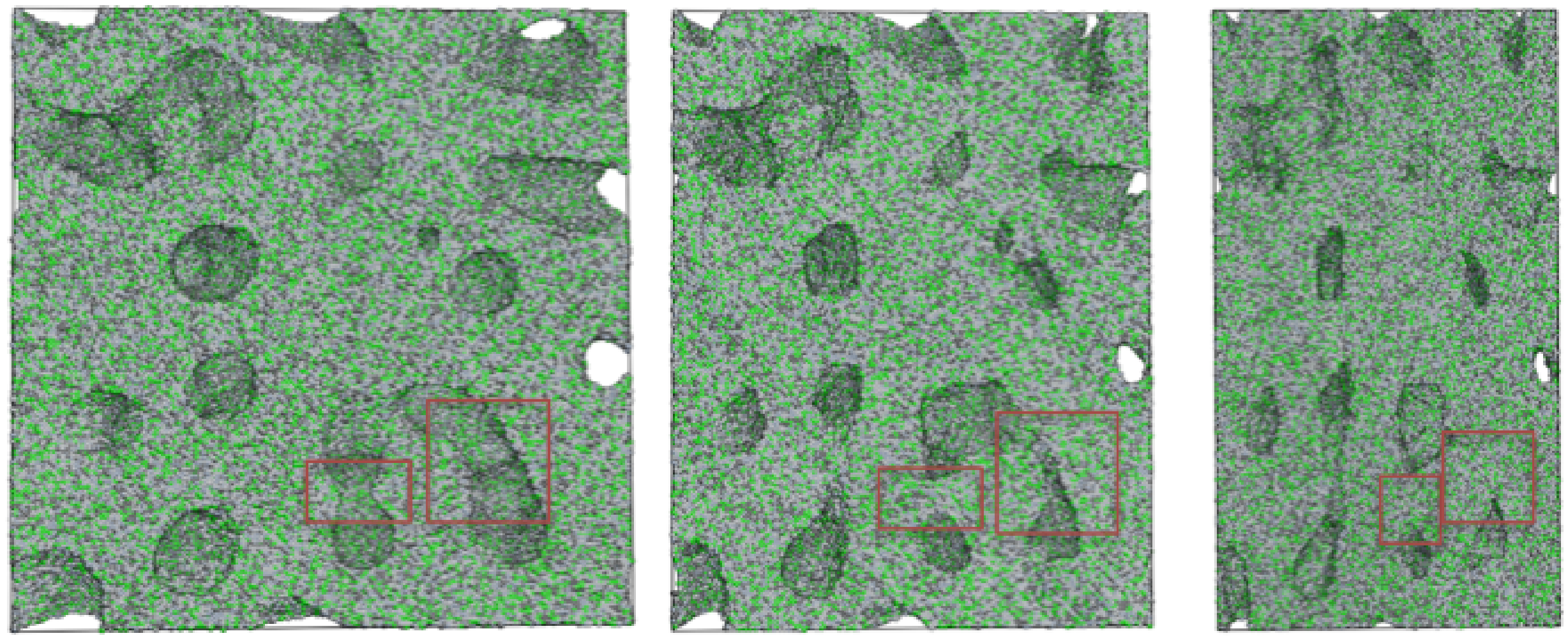

**Figure S10.** Simulation boxes showing the 67.2% relative density 7.33 nm $Al_{0.1}CoCrFeNi$ nanoporous system under compression at 298 K. The red boxes show the locations of pore sites being reduced and ligaments connecting with strains of 0% (a) / 25% (b) / 50% (c).

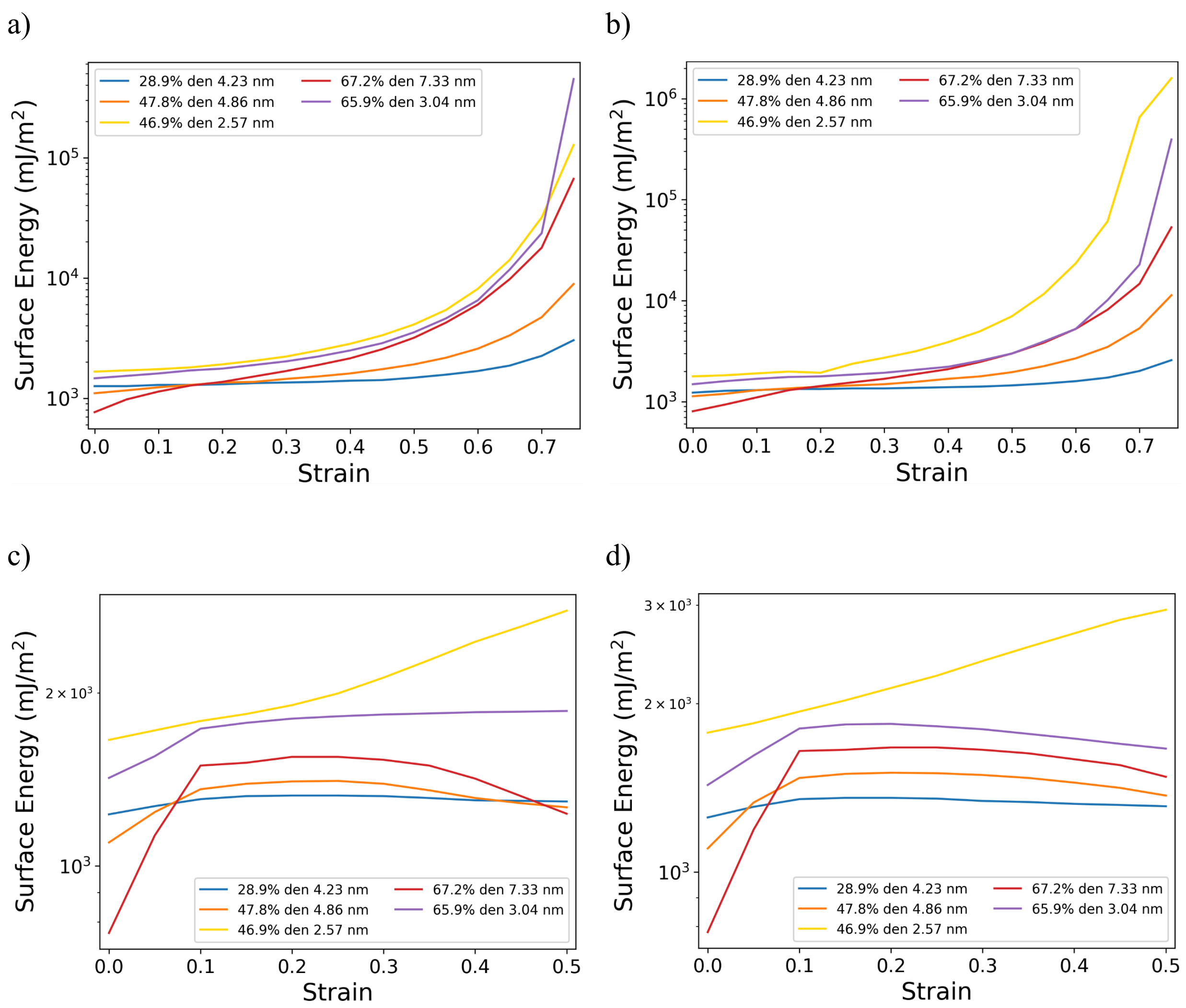

**Figure S11.** Surface energy vs strain of nanoporous $Al_{0.1}$CoCrFeNi for compression at 600 K (a) / 1273 K (b) and tension at 600 K (c) / 1273 K (d).

**Supplementary Table 1.** Yield strengths of nanoporous NbMoTaW, 28.0% relative density with different tension strain rates.

| Strain | Ultimate Strength (GPa) |
| --- | --- |
| $5 \times 10^{8}$ | 0.47 |
| $1 \times 10^{9}$ | 0.49 |
| $5 \times 10^{9}$ | 0.51 |
| $1 \times 10^{10}$ | 0.61 |